\documentclass[aps,prb,showpacs,amsmath,twocolumn]{revtex4-1}

\usepackage{amsmath}
\usepackage{amssymb}
\usepackage{graphicx}
\usepackage{bbm}

\newcommand{\vareps}{\varepsilon}

\begin{document}

\title{Intrinsic and substrate induced spin-orbit interaction in chirally stacked 
trilayer graphene}

\author{Andor Korm\'anyos}
\thanks{e-mail: andor.kormanyos@uni-konstanz.de}
\affiliation{Department of Physics, University of Konstanz, D-78464 Konstanz, Germany
}

\author{Guido Burkard}
\affiliation{Department of Physics, University of Konstanz, D-78464 Konstanz, Germany
}

%\wideabs{

\begin{abstract}
We present a combined group-theoretical and  tight-binding approach to calculate the 
intrinsic spin-orbit coupling (SOC) in ABC stacked trilayer graphene. 
We find that compared to monolayer graphene 
(S. Konschuh, M. Gmitra, and J. Fabian [Phys. Rev. B 82, 245412 (2010)])\cite{jfabian-mono-TB}, 
a larger set of $d$ orbitals (in particular the $d_{z^2}$ orbital)
needs to be taken into account. 
We also consider the intrinsic SOC in bilayer graphene, because the comparison between our tight-binding  
 bilayer results and the density functional computations of (Ref.~\onlinecite{jfabian-bilayer} allows us 
 to estimate  the values of the trilayer SOC parameters as well.
We also discuss the situation when a substrate or adatoms induce strong SOC in only one of the  layers 
of bilayer or ABC trilayer graphene. Both for the case of  intrinsic and    externally induced SOC we 
derive effective Hamiltonians which describe the low-energy spin-orbit physics.
We find that at the $K$ point of the Brillouin zone
the effect of Bychkov-Rashba type SOC is suppressed in bilayer and ABC trilayer graphene  compared to 
monolayer graphene. 

\end{abstract} 


\maketitle

\section{Introduction}
\label{sec:intro}

The low-energy properties of multilayer graphene\cite{graphene-review} 
depend crucially on the stacking order of the constituent graphene 
layers\cite{multilayer-guinea,lu,luc-hernard,aoki,partoens,rubio,mccann-abctrilayer,mccann-abatrilayer,koshino,macdonald-1,
mak,hung-lui,nphys-1,nphys-2,nphys-3,nphys-4}.
In the case of  trilayer graphene, there are  two stable stacking orders: (i) ABA or Bernard stacking 
and  (ii) ABC or chiral stacking. Recent  advances in sample fabrication methods have resulted in   
high-quality trilayer samples which can be used  to probe many of the theoretical 
predictions\cite{mak,hung-lui,nphys-1,nphys-2,nphys-3,nphys-4,khodkov}. 
 ABC stacked trilayer graphene appears to be particularly exciting because it  is expected to host a wealth of 
interesting phenomena, such as chiral quasiparticles with Berry phase $3\pi$\cite{mccann-abctrilayer}, 
a Lifshitz transition of electronic bands due to trigonal warping\cite{mccann-abctrilayer,macdonald-1}, 
band-gap opening in an external electric 
field\cite{multilayer-guinea,lu,aoki,mccann-abctrilayer,koshino,macdonald-1,avetisyan,abc-dft-xiao,nphys-1,khodkov}, 
and broken symmetry phases at low electron densities\cite{macdonald-2,macdonald-3,barlas,cvetkovic}, to name a few. 

Although there are a number of 
theoretical\cite{multilayer-guinea,aoki,lu,luc-hernard,partoens,mccann-abctrilayer,mccann-abatrilayer,koshino,
rubio,macdonald-1,avetisyan,abc-dft-xiao,kumar,castro} 
and experimental\cite{ensslin,nnanotech,mak,kumar,hung-lui,nphys-1,nphys-2,nphys-3,nphys-4} studies on the 
electronic properties of ABA and ABC stacked trilayer graphene, 
the spin-orbit coupling (SOC) in these systems has received much less attention.
ABA  trilayer graphene was considered in Ref.~\onlinecite{mccann-spinorbit} within a framework 
of an effective low-energy theory, whereas the case ABC stacking was only  briefly mentioned 
in Ref.~\onlinecite{guinea-so}. The understanding of spin-orbit interaction  would  be important 
to study other interesting and  experimentally relevant 
phenomena such as  spin relaxation\cite{ertler,mccann-monolayerspinorb,ochoa,diez}, 
weak-localization\cite{mccann-monolayerspinorb} or even spin-Hall effect\cite{kane-mele} in trilayer graphene. 
The recent report of Ref.~\onlinecite{khodkov} on the fabrication of high mobility 
double gated ABC trilayer graphene may open very promising new avenues for trilayer graphene spintronics as well, 
similarly to the monolayer case where highly efficient spin transport has recently been reported\cite{fert}, but 
with the additional advantage that external gates can open a band gap in ABC trilayer graphene.

In this paper we aim to make the first steps towards a detailed understanding of  the spin-orbit coupling (SOC) 
in chirally stacked trilayer graphene. We start by investigating the case when the system has inversion
symmetry, i.e. in the absence of external electric fields, adatoms or a substrate. 
This is the case of \emph{intrinsic} SOC. 
The intrinsic SOC  opens a band-gap at the band-degeneracy points without introducing  spin polarization.  
Previous \emph{ab initio} calculations on monolayer\cite{jfabian-so-numeric,abdelouahed} and 
bilayer\cite{jfabian-bilayer} graphene %have shown
provided   strong evidence that the key to the understanding the 
SOC in \emph{flat}  graphene systems is to   take into account  the (nominally unoccupied)
$d$ orbitals in the description of  electronic bands.
Here we take the same view and by 
generalizing the work of Ref.~\onlinecite{jfabian-mono-TB} 
 derive the intrinsic SOC Hamiltonian for ABC trilayer graphene.  
It turns out that the most important $d$ orbitals to take into account are 
the $d_{xz}$, $d_{yz}$ and $d_{z^2}$ orbitals. 
While the former two have been considered in  Ref.~\onlinecite{jfabian-mono-TB} 
in the context of  monolayer 
graphene, the latter one is important to understand the SOC in AB and ABC graphene.
We obtain explicit expressions for the SOC constants
in terms of Slater-Koster\cite{slater-koster} hopping parameters. We also 
 rederive the intrinsic SO Hamiltonian for bilayer graphene\cite{guinea-so,jfabian-bilayer}. 
Through the comparison of our trilayer and bilayer analytical results with  the recent \emph{ab initio}  
calculations of Ref.~\onlinecite{jfabian-bilayer}
we are able to make predictions for the actual values of the SOC parameters in ABC trilayer graphene.  
The theory involves electronic bands which are far from the Fermi energy $E_{\rm F}$ but are coupled to the
physically important low-energy bands  close to  $E_{\rm F}$ 
and hence complicate the description of the electronic properties.
Therefore, we derive an effective low-energy Hamiltonian and calculate its spectrum. 
This helps us to understand how SOC lifts certain degeneracies of the electronic bands.   

Generally speaking, due to the low atomic  number of  carbon, the  intrinsic SOC in single and multilayer graphene is weak 
(according to our prediction, the  SOC parameters are of  the order of  $10\,\mu{\rm eV}$ in ABC trilayer, the 
same as in monolayer\cite{boettger,jfabian-so-numeric,jfabian-mono-TB,abdelouahed} and bilayer\cite{jfabian-bilayer} graphene). 
Recently however, there have been  exciting theoretical proposals to enhance the strength of SOC in monolayer graphene 
and hence, e.g., make the quantum spin Hall state\cite{kane-mele} observable.  
These proposals suggest   deposition of indium or thallium atoms\cite{graphene-adatom} 
or to bring  graphene into proximity with topological insulators\cite{proximity-spin-orbit}. 
Indeed, very recently the combined experimental and theoretical work of  Ref.~\onlinecite{varykhalov-nature} has 
provided evidence of a large ($10-100\, {\rm meV}$) spin-orbit gap in monolayer graphene on nickel substrate 
with intercalated gold atoms. 
%Here the hybridization of the carbon atoms with the gold atoms leads to a Rashba type\cite{rashba} spin splitting.
Motivated by these studies we also discuss what might be a minimal model to describe the case where the 
the SOC is strongly enhanced in only one of the  layers of  bilayer and ABC trilayer graphene. 

Our work is organized as follows. In Sect.~\ref{sec:TB-model} we  present the tight-binding (TB) model of 
ABC-stacked graphene and introduce certain notations that we will be using in subsequent sections. 
In Sect.~\ref{sec:intrinsic-soc}, employing group-theoretical considerations and the Slater-Koster\cite{slater-koster} 
(SK) parametrization of transfer integrals, we derive the SOC Hamiltonian in atomistic approximation at the $K$ 
point of the Brillouin zone. 
We repeat this calculation for bilayer graphene in Sect.~\ref{sec:bilayer} so that in Sect.~\ref{sec:SO-ABC-SL} 
we can make predictions for the actual values of the SOC parameters. Next, in Section \ref{sec:eff_so_ham},
using $\mathbf{k}\cdotp\mathbf{p}$ 
theory and the Schrieffer-Wolff transformation\cite{winkler-book,loss-schrieffer-wolff}, we derive an effective low-energy 
SOC Hamiltonian which is valid  %not only at the $K$ ($K'$) point but also 
for wave-vectors around the  $K$ ($K'$) point. 
Finally, in Sect.~\ref{sec:induced-soc}, we consider the  case when SOC is enhanced in one of the graphene layers 
with respect to the other(s).

\section{Tight-binding model}
\label{sec:TB-model}

\begin{figure}[ht]
\includegraphics[scale=0.5]{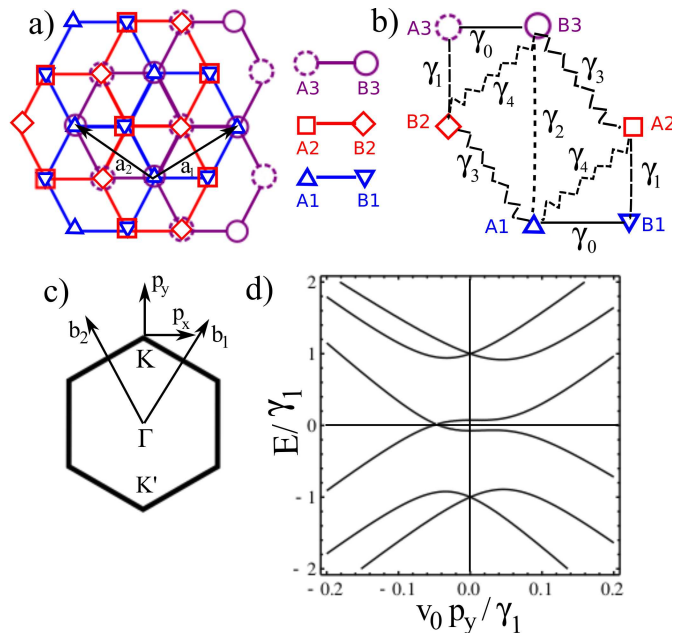}
\caption{Lattice and band structure of ABC trilayer graphene. 
         a) Lattice structure, where atoms on different layers are indicated with different symbols. 
            $\mathbf{a}_1$ and $\mathbf{a}_2$ are the two lattice vectors.
         b) Schematic side view of the unit cell with the most important hopping amplitudes.
         c) Schematic of the Brillouin zone with reciprocal lattice vectors $\mathbf{b}_1$, $\mathbf{b}_2$, 
          the high-symmetry points $\Gamma$, $K$ and $K'$ and the 
            {momentum $\mathbf{p}=(p_x,p_y)$ measured from $K$}.
         d) Schematic of the band structure at the $K$ point of the Brillouin zone. {The velocity $v_0$ is 
            given by $v_0=\sqrt{3}/2 a \gamma_0/\hbar$, where $a$  is the lattice constant: $a=|\mathbf{a}_{1}|$}. }
 \label{fig1}
\end{figure}

The basic electronic properties of ABC trilayer are well captured by the effective mass model which 
is derived assuming one $p_z$ type atomic orbital per carbon atom. This model has been 
discussed in detail in Refs.~\onlinecite{mccann-abctrilayer} and \onlinecite{macdonald-1},; 
therefore we  give only a very short introduction here (see also Fig.~\ref{fig1}).
There are six carbon atoms 
in the unit cell of ABC trilayer graphene, usually denoted by $A1$, $B1$, $A2$, $B2$, $A3$, $B3$, where 
$A$ and $B$ denote the sublattices and $j=1,2,3$ is the layer index. The parameters appearing in the 
effective model are: $\gamma_0$ for the intra-layer $Aj-Bj$ nearest-neighbour hopping, 
$\gamma_1$  for the interlayer hopping between sites $B1-A2$ and $B2-A3$, $\gamma_3$ 
($\gamma_4$) describes weaker nearest-layer hopping between atoms 
belonging to different (the same) sublattice, 
and finally $\gamma_2$ denotes the direct hopping between sites
$A1$ and $B3$ that lie on the same vertical linein the outer layers $1$ and $3$. These hoppings can be 
obtained by e.g. fitting the numerically calculated band structure with a TB model\cite{rubio,macdonald-1}.
The six $p_z$ orbitals in the unit cell give rise to six electronic bands\cite{mccann-abctrilayer,macdonald-1}. 
As  shown in Fig.\ref{fig1}(d), 
at the $K$ ($K'$) point of the Brillouin zone (BZ) two of these bands lie 
close to the Fermi energy $E_{\rm F}=0$ and  we will refer to them as "low-energy`` states. 
In addition, there are four ''split-off`` states far from $E_{\rm F}$ at energies $E=\pm \gamma_1$.

To obtain the intrinsic SOC Hamiltonian of ABC trilayer graphene we generalize the main idea of 
Ref.~\onlinecite{jfabian-mono-TB} where the SOC of monolayer graphene was discussed.
Using group-theoretical considerations and  density functional theory (DFT) calculations it was shown 
 in Ref.~\onlinecite{jfabian-mono-TB} that in the case of monolayer graphene  the main contribution 
to the intrinsic SOC  comes from the admixture of $p_z$ orbitals with some of the (nominally unoccupied)
$d$ orbitals,  namely,  with the  $d_{xz}$ and $d_{yz}$ orbitals. 
The other $d$ orbitals, $d_{z^2}$, $d_{xy}$ and 
$d_{x^2-y^2}$ play no role due to the fact that they are symmetric with respect to the mirror 
reflection  $\sigma_{h}: (x,y,z)\rightarrow (x,y,-z)$  to the plane of the graphene layer, 
whereas $p_z$ is antisymmetric. 
The importance of the $d$ orbitals for the understanding of SOC in monolayer graphene
was also pointed out in Ref.~\onlinecite{abdelouahed}. 

The symmetry group of ABC stacked trilayer graphene  is
$R\bar{3}m$ ($D_{3d}$) which does not contain the mirror reflection $\sigma_{h}$.  Therefore, 
from a symmetry point of view, in an approach similar to Ref.~\onlinecite{jfabian-mono-TB}, 
all $d$ orbitals  need to be taken into account.  
To derive a TB model one should therefore use as  basis set the Bloch wavefunctions
\begin{equation}
 \Psi^{\alpha j}_{\beta}(\mathbf{r},\mathbf{k})=\frac{1}{\sqrt{N}} \sum_{n} e^{i \mathbf{k}\cdotp (\mathbf{R}_n+\mathbf{t}_{\alpha})}
 \Phi_{\beta}^{j}(\mathbf{r}-(\mathbf{R}_n+\mathbf{t}_{\alpha})), 
 \label{Bloch-functs}
\end{equation}
where the wavevector $\mathbf{k}$ is measured from the $\Gamma$ point of the BZ (see Fig.~\ref{fig1}),  
$\alpha j=\{A1,A2,A3,B1,B2,B3\}$ is a composite index for the sublattice $\alpha=\{A$, $B\} $ and layer $j=\{1,2, 3\}$ 
indices and $\Phi_{\beta}^{j}$ denotes the atomic orbitals of type  $\beta=\{p_z, d_{z^2}, d_{xz}, d_{yz}, d_{xy}, d_{x^2-y^2}\}$ 
in layer $j$. 
The summation runs over all Bravais lattice vectors $\mathbf{R}_n$, whereas the vectors $\mathbf{t}_{\alpha}$ give the
position of atom $\alpha$ in the two-dimensional unit cell. We use a coordinate system where the primive lattice vectors
are $\mathbf{a}_1=\frac{a}{2} (\sqrt{3},1)$ and $\mathbf{a}_2=\frac{a}{2} (-\sqrt{3},1)$, the positions of the atoms 
in the unit cell are $\mathbf{t}_{A1}=\mathbf{t}_{B3}=(0,0)$,  
$\mathbf{t}_{A2}=\mathbf{t}_{B1}=\frac{a}{2}\left(-\frac{1}{\sqrt{3}}, 1\right)$ and
 $\mathbf{t}_{A3}=\mathbf{t}_{B2}=\frac{a}{2}\left(\frac{1}{\sqrt{3}}, 1\right)$ where $a=2.46 \AA{}$ is the lattice constant. 
 The $K$ and $K'$ points of the Brillouin zone, which are important for the low energy physics discussed in this paper, 
 can be found at $K=(0,\frac{4 \pi}{3 a})$,  $K'=(0,-\frac{4 \pi}{3 a})$.
 
 The symmetry group of the lattice contains  threefold
 rotations by $\pm {2 \pi}/{3}$, about an axis perpendicular to the graphene layers. Since the atomic orbitals 
 $d_{xz}, d_{yz}, d_{xy}, d_{x^2-y^2}$  themselves do not possess this symmetry, instead of  
 $ \Psi^{\alpha j}_{\beta}(\mathbf{r},\mathbf{k})$  given in Eq.~(\ref{Bloch-functs})  
 we  will use Bloch states which depend on $\frac{1}{\sqrt{2}}(d_{xz}\pm i d_{yz})\sim \mp Y_{2}^{\pm 1}$, 
$\frac{1}{\sqrt{2}}(d_{x^2-y^2}\pm i d_{xy})\sim  Y_{2}^{\pm 2}$ (rotating orbitals), where $Y_{l}^{m}$ are spherical harmonics. 
Taking into account that $p_z\sim Y_{1}^{0}$ and $d_{z^2}\sim Y_{2}^{0}$, the Bloch states we use as basis will be denoted by
$\Psi^{\alpha j}_{l,m_l}(\mathbf{r},\mathbf{k})$, where $l=1,2$ and $m_1=0$, whereas $m_2$ can take all allowed
values $m_2=-2\dots 2$. 
Often, we will need  a linear combination of two of these
basis functions where both of the basis functions have the same quantum number $l$  but 
one of them is centered on an $A$ type atom  and the other one is on a $B$ type atom,  
 e.g. $\frac{1}{\sqrt{2}}[\Psi^{A1}_{1,0}(\mathbf{r},\mathbf{k})-\Psi^{B3}_{1,0}(\mathbf{r},\mathbf{k})]$.
As a shorthand notation, we will denote the symmetric combination of two such basis functions 
%(i.e. with "+`` sign between the constituent functions) 
by $\Psi^{j,j'}_{l,m,m'}(\mathbf{r},\mathbf{k})=
\frac{1}{\sqrt{2}}[\Psi^{A j}_{l,m}(\mathbf{r},\mathbf{k})+\Psi^{B j'}_{l,m'}(\mathbf{r},\mathbf{k})]
$ 
and the antisymmetric one with  
$
\overline{\Psi^{j,j'}_{l,m,m'}(\mathbf{r},\mathbf{k})} = 
\frac{1}{\sqrt{2}}[\Psi^{A j}_{l,m}(\mathbf{r},\mathbf{k})-\Psi^{B j'}_{l,m'}(\mathbf{r},\mathbf{k})].
$ 
The first upper index $j$ in $\Psi^{j,j'}_{l,m,m'}$ always denotes  the layer index of the atomic 
orbital centered on the $A$ type atom, the second upper index $j'$ is the layer index for  
the orbital centered on  the $B$ type atom, the first lower index $l$ is the common angular momentum quantum number, 
and finally, the second and the third lower indices $m$, $m'$ give the magnetic quantum number in the same manner as
the upper indices give the layer index. To lighten  the notation, we will usually suppress the
dependence of the Bloch functions on $(\mathbf{r},\mathbf{k})$ and  use 
the bra-ket notation, e.g. $|\Psi^{1,3}_{1,0,0} \rangle$, $|\overline{\Psi^{1,3}_{1,0,0} }\rangle$.

The derivation of the spin-orbit Hamiltonian proceeds in the same spirit as in 
Ref.~\onlinecite{jfabian-mono-TB}: (i) First, we neglect the spin degree of freedom.  
Using the  Slater-Koster  parametrization to describe the hopping integrals  
$\langle \Psi^{j,j'}_{l,m,m'}|\hat{\mathcal{H}}| \Psi^{j'',j'''}_{l',m'',m'''}\rangle$ 
($\hat{\mathcal{H}}$ is the single particle Hamiltonian of the system)
at a high symmetry point (the $K$ point) of the Brillouin zone 
and group-theoretical considerations we obtain  
certain effective Bloch wavefunctions  which  comprise  $p_z$ and $d$ atomic orbitals centered on 
different atoms;
(ii) using these effective wavefunctions we calculate the matrix elements of the 
 spin-orbit Hamiltonian in atomic approximation, and 
 (iii) employing the $\mathbf{k}\cdotp\mathbf{p}$ theory we obtain the bands around the
 $K$ point and then we derive an effective low-energy Hamiltonian.

\section{Intrinsic SOC}
\label{sec:intrinsic-soc}

If, in addition to the $p_z$ orbitals,  we include also the $d$ orbitals into our basis, 
there will be six basis functions 
$|\Psi_{l,m}^{\alpha j}\rangle$ centered on each of the six carbon atoms in the unit cell and hence 
the TB Hamiltonian ${H}_{ABC}$, which is straightforward to calculate in the SK parametrization, 
is a $36\times 36$ matrix. By e.g. numerically diagonalizing this matrix one would find that 
the $p_z$ orbitals hybridize with some of the $d$ orbitals and one could see how the 
low energy and the split-off states, obtained in the first
instance by neglecting the $d$ orbitals, are modified. 
According to band theory  each state at the $K$ point 
should belong to  one of the irreducible representations of the  small group of the $K$ 
point\cite{dresselhaus-book}, which is $32$ ($D_3$) in this case. 
This group has two one-dimensional irreducible representation, 
denoted by $\Gamma_{A_1}$ and $\Gamma_{A_2}$ respectively, 
and a two-dimensional one 
denoted by $\Gamma_{E}$ (see Appendix \ref{sec:irrep-basis-funct}). 
The matrix elements of  ${H}_{ABC}$ between basis states 
corresponding to different irreducible representations of $D_3$ 
are zero\cite{dresselhaus-book}. 
In other words, ${H}_{ABC}$ can be block-diagonalized by 
choosing suitable  linear combinations of the basis functions 
$|\Psi_{l,m}^{\alpha j}\rangle$ such that the  new basis functions transform as the irreducible representations
of the group $D_3$ because the hybridization between $p_z$ and $d$ orbitals will
 preserve the symmetry properties. 
A group-theoretical analysis of the problem  shows that in a suitable basis  ${H}_{ABC}$ is
block-diagonal having (i) two $6\times 6$ blocks which we denote by $H_{\Gamma_{A_1}}$ 
and $H_{\Gamma_{A_2}}$, they  correspond to basis states with $\Gamma_{A_1}$ and
$\Gamma_{A_2}$ symmetry, and (ii) there is  one $24\times 24$ block  ${H}_{\Gamma_{E}}$
corresponding to states with $\Gamma_{E}$ symmetry. 
(The basis vectors with $\Gamma_{A_1}$, $\Gamma_{A_2}$ and $\Gamma_{E}$ symmetries are listed in 
Appendix \ref{sec:irrep-basis-funct}, Table \ref{tbl:symmetry-basis-full-trilayer}).
As a concrete example we will consider  $H_{\Gamma_{A_1}}$ and discuss how 
one can extract an effective orbital in which  $p_z$ atomic orbitals with large weight and
$d$ orbitals with small weight are admixed.  The calculation for $H_{\Gamma_{A_2}}$ 
and ${H}_{\Gamma_{E}}$ cases is analogous and will be presented only briefly.  

The basis states transforming as the irreducible representation $\Gamma_{A_1}$ are 
$|\overline{\Psi_{1,0,0}^{1,3}}\rangle$, $|\Psi_{2,1,-1}^{3,1}\rangle$, $|\Psi_{2,0,0}^{1,3}\rangle$, 
$|\Psi_{2,-2,2}^{3,1}\rangle$, $|\Psi_{2,-1,1}^{2,2}\rangle$ and $|\Psi_{2,2,-2}^{2,2}\rangle$.
The  TB Hamiltonian $H_{\Gamma_{A_1}}$  can be further divided into $3\times 3$  blocks: 
%each block being a $3\times 3$ matrix:
\begin{equation}
 H_{\Gamma_{A_1}}=\left(
\begin{array}{cc}
 H_{pd}^{A_1} & W^{A_1} \\
 (W^{A_1})^{\dagger} & H_{dd}^{A_1}
\end{array}
\right).
\label{A1_TB_Ham}
\end{equation}
Explicitly, the upper left block $H_{pd}^{A_1}$, corresponding to the basis states
$|\overline{\Psi_{1,0,0}^{1,3}}\rangle, |\Psi_{2,1,-1}^{3,1}\rangle$ and $|\Psi_{2,0,0}^{1,3}\rangle$, reads 
\begin{eqnarray}
 H_{pd}^{A_1}=\left(
\begin{array}{ccc}
 \vareps_p-\gamma_2 & \frac{3}{\sqrt{2}} V_{p d \pi} & V_{p d \sigma}^{A1,B3} \\
 \frac{3}{\sqrt{2}} V_{p d \pi} & \vareps_d  &  0 \\
  V_{p d \sigma}^{A1,B3} & 0 & \vareps_d+V_{d d \sigma}^{A1,B3}
 \end{array}
\right),
\end{eqnarray}
where the upper indices $A1$, $B3$ on the SK parameters indicate the atomic sites between which the hopping takes place. 
The parameter $V_{p d \pi}$ describes hopping between $A$ and $B$ type atoms  within the same graphene layer and
we assume that its value is the same in all three layers. 
The matrix elements in $W^{A_1}$ of Hamiltonian (\ref{A1_TB_Ham}) are either zero\cite{zeros} 
or describe skew hoppings between the 
$p_z$ and $d$ orbitals located on different atoms. We assume that these skew hoppings are much smaller than 
both the vertical hopping $V_{p d \sigma}^{A1,B3}$ and  $V_{p d \pi}$. 
This is not a crucial assumption  and   the neglected skew-hoppings can be taken into account in a 
straightforward manner. However, it simplifies  the lengthy algebra that follows and we believe it yields 
qualitatively correct results (see Section~\ref{sec:bilayer}). 
With $W^{A_1} \approx 0$ we see that $H_{dd}^{A_1}$ (corresponding 
to the basis functions $|\Psi_{2,-2,2}^{3,1}\rangle$, $|\Psi_{2,-1,1}^{2,2}\rangle$ and $|\Psi_{2,2,-2}^{2,2}\rangle$) 
is  decoupled  from $H_{pd}$ and that this latter matrix describes  the 
hybridization between the  $p_z$ orbital based basis vector $|\overline{\Psi_{1,0,0}^{1,3}}\rangle$ and the 
basis vectors $|\Psi_{2,1,-1}^{3,1}\rangle$, $|\Psi_{2,0,0}^{1,3}\rangle$ involving $d_{xz}$, $d_{yz}$ and $d_{z^2}$  orbitals. 
By diagonalizing $H_{pd}^{A_1}$  one could find out how the energy $\vareps_{A_1}=\vareps_p-\gamma_2$ 
of one of the low-energy states is modified by the $d$ orbitals. 
The secular equation $\det(H_{pd}^{A_1}- \vareps I )=0$ leads to a cubic equation in $\vareps$ and the solutions can only
be expressed using the Cardano formula. 
Instead, we next perform a Schrieffer-Wolff transformation (L\"owdin partitioning) 
$\tilde{H}_{pd}^{A_1} = e^{-S}\, H_{pd}^{A_1}\, e^{S}$ 
%where $e^{-S}$ is a unitary matrix, 
to approximately block-diagonalize $H_{pd}^{A_1}$ into a $1 \times 1$ and a $2 \times 2$ block 
by eliminating the matrix elements between 
$|\overline{\Psi_{1,0,0}^{1,3}}\rangle$ on one hand and  
$|\Psi_{2,1,-1}^{3,1}\rangle$, $|\Psi_{2,0,0}^{1,3}\rangle$ on the other hand. 
(A detailed discussion of this method can be found in e.g. Refs.~\onlinecite{winkler-book,loss-schrieffer-wolff}.) 
The matrix $S$  is anti-Hermitian: $S^{\dagger}=-S$ and only its nondiagonal blocks $S_{pd}$ and $-S_{pd}^{\dagger}$ are non-zero. 
In first order\cite{winkler-book} of the coupling matrix elements $V_{p d \pi}$ and $V_{p d \sigma}^{A1,B3}$ 
one finds that
\begin{equation}
 S^{(1)}_{pd} =-\left(\frac{\bar{V}_{p d \pi}}{\delta\vareps_{pd}-\gamma_2}, 
 \frac{V_{p d \sigma}^{A1,B3}}{\delta\vareps_{pd}-\tilde{\gamma}_2}\right).
 \label{S-approx}
\end{equation}
where $\bar{V}_{p d \pi}=\frac{3}{\sqrt{2}} V_{p d \pi}$, $\delta \vareps_{pd}=\vareps_p-\vareps_d$ and 
$\tilde{\gamma}_2=\gamma_2+V_{d d \sigma}^{A1, B3}$.  The $1\times 1$ block of $\tilde{H}_{pd}^{A_1}$ 
reads 
$
\tilde{\vareps}_{A_1}=\vareps_p-\gamma_2+ {\bar{V}_{p d  \pi}^2}/{(\delta\vareps_{pd}-\gamma_2)} +  
{(V_{p d \sigma}^{A1,B3})^2}/{(\delta\vareps_{pd}-\tilde{\gamma}_2)}
$,
this means that the energy $\vareps_{A_1}$ of the low energy state is shifted by 
$
{\bar{V}_{p d  \pi}^2}/{(\delta\vareps_{pd}-\gamma_2)} +  
{(V_{p d \sigma}^{A1,B3})^2}/{(\delta\vareps_{pd}-\tilde{\gamma}_2)}.
$ 
While the Schrieffer-Wolff  transformation is usually used to obtain effective Hamiltonians, 
 one can  also obtain the new basis in which $\tilde{H}_{pd}^{A_1}$ is blockdiagonal. 
Making the approximation $e^{-S}\approx 1-S $ (see Ref.~\onlinecite{S-approx}) we find that the purely 
$p_z$-like state  $|\overline{\Psi_{1,0,0}^{1,3}}\rangle$ is transformed into 
\begin{equation}
 |\Psi_{\Gamma_{A_1}}^{p_z}\rangle= |\overline{\Psi_{1,0,0}^{1,3}}\rangle + 
                               \frac{\bar{V}_{p d \pi}}{\delta \vareps_{pd}-\gamma_2}|\Psi_{2,1,-1}^{3,1}\rangle +
                               \frac{V_{p d \sigma}^{A1,B3}}{\delta\vareps_{pd}-\tilde{\gamma}_2} |\Psi_{2,0,0}^{1,3}\rangle,                               
\end{equation}
i.e. it is admixed with  two other basis vectors containing $d_{xz}, d_{yz}$ and $d_{z^2}$ orbitals. 
Since $|\Psi_{\Gamma_{A_1}}^{p_z}\rangle$ corresponds to a $1 \times 1$ and hence diagonal block of 
$\tilde{H}_{pd}$,  it is an approximate eigenvector of $H_{pd}^{A_1}$ with 
energy $\tilde{\vareps}_{A_1}$. The upper index $p_z$ in $|\Psi_{\Gamma_{A_1}}^{p_z}\rangle$ is meant to indicate
that in this state $p_z$ orbitals have the largest weight. 
There are two other states with $\Gamma_{A_1}$ symmetry which could be obtained by diagonalizing the 
the remaining $2 \times 2$ block of $\tilde{H}_{pd}^{A_1}$. In these states 
$|\Psi_{2,1,-1}^{3,1}\rangle$ and $|\Psi_{2,0,0}^{1,3}\rangle$  would have large weight. 
They  are however far remote in
energy from   $|\Psi_{\Gamma_{A_1}}^{p_z}\rangle$ and therefore play no role in our further considerations. 
The situation will be similar in the case of the two other irreducible representations, 
$\Gamma_{A_2}$ and $\Gamma_{E}$, therefore we 
will suppress the upper index $p_z$  henceforth in the notation 
of the physically important approximate eigenstates. 

We now briefly discuss the symmetry classes $\Gamma_{A_2}$ and $\Gamma_{E}$. The calculation for the other 
$6 \times 6$ block of $H_{ABC }$  with  $\Gamma_{A_2}$ symmetry is completely analogous to the  $\Gamma_{A_1}$ case, 
the resulting approximate eigenvector, $|\Psi_{\Gamma_{A_2}}^{}\rangle$ is shown in the left 
column of Table \ref{tbl:SlatKost-SOC}.
Its energy, apart from the  shift due to the $d$ orbitals, 
which will be neglected, is $\vareps_{A_2}=\vareps_p+\gamma_2$.

The matrix block corresponding to states with $\Gamma_{E}$ symmetry can  be written as 
\begin{equation}
  H_{\Gamma_{E}}=\left(
\begin{array}{cc}
 H_{pp}^{E} & W^{E}_{pd} \\
 (W^{E}_{pd})^{\dagger} & H_{dd}^{E}
\end{array}
\right).
\label{E_TB_Ham}
\end{equation}
Here the $4 \times 4 $ block  $H_{pp}^{E}$ contains the matrix elements between the basis vectors 
$|{\Psi_{1,0,0}^{2,1}}\rangle$, $|{\Psi_{1,0,0}^{3,2}}\rangle$, $|\overline{\Psi_{1,0,0}^{2,1}}\rangle$,
$|\overline{\Psi_{1,0,0}^{3,2}}\rangle$, 
the $4 \times 20$ block  $W_{pd}^{E}$ is the coupling  matrix between the above shown $p_z$ orbital 
based basis vectors and the $d$ orbitals based basis vectors. (The full set of basis vectors  for each of the 
three symmetry classes is listed in Appendix \ref{sec:irrep-basis-funct}.)
A direct calculation shows that $H_{pp}$ is  a diagonal matrix with 
$\vareps_p+\gamma_1,\vareps_p+\gamma_1,\vareps_p-\gamma_1,\vareps_p-\gamma_1$ entries. 
Earlier we have referred to these states as ''split-off`` states. Since $H_{pp}^{E}$ is diagonal,  its approximate eigenvectors 
$ |\Psi_{\Gamma_{E_{1,1}}}^{}\rangle$, $|\Psi_{\Gamma_{E_{1,2}}}^{}\rangle$, 
$|\Psi_{\Gamma_{E_{2,1}}}^{}\rangle$ and $|\Psi_{\Gamma_{E_{2,2}}}^{}\rangle$, which are listed in Table \ref{tbl:SlatKost-SOC}, 
can be obtained in exactly the same way as $|\Psi_{\Gamma_{A_{1}}}^{}\rangle$.

We will refer to  the  basis  formed from the physically important approximate eigenvectors 
$
\{ 
|\Psi_{\Gamma_{A_{1}}}\rangle, |\Psi_{\Gamma_{A_{2}}}\rangle, |\Psi_{\Gamma_{E_{1,1}}}\rangle,
|\Psi_{\Gamma_{E_{1,2}}}^{}\rangle, |\Psi_{\Gamma_{E_{2,1}}}^{}\rangle, |\Psi_{\Gamma_{E_{2,2}}}^{}\rangle
\}
$ 
 as  the ''symmetry basis`` henceforth. (The symmetry basis for the $K'$ point can be obtained by complex-conjugation.) 
Looking at these basis vectors we see  that in contrast to monolayer graphene\cite{jfabian-mono-TB}, 
where only $d_{xz}$ and $d_{yz}$ orbitals hybridize with the $p_z$ orbital, here also the $d_{z^2}$ orbitals are admixed. 
As it will be shown below, the admixture of $d_{z^2}$ orbitals is crucial to 
obtain the non-diagonal  elements of the SOC Hamiltonian.

We can now proceed to calculate the SOC Hamiltonian. This can be done in the atomic approximation, whereby the spin-orbit 
interaction is described by the Hamiltonian
\begin{equation}
 \hat{H}_{SO}^{atomic}=\frac{\hbar}{4 m_e^2 c^2} \frac{1}{r} \frac{d V(r)}{d r} \,\mathbf{L} \cdotp \mathbf{S}
\end{equation}
Here $V(r)$ is the spherically symmetric atomic potential, $\mathbf{L}$ is the  angular momentum operator and 
$\mathbf{S}=(S_x,S_y)$ is a vector of spin Pauli matrices $S_x,\,S_y$ (with eigenvalues $\pm 1$). 
Introducing the spinful symmetry basis functions by $| \Psi_{\mu}\rangle \rightarrow | \Psi_{\mu}\otimes s \rangle$, 
where $s=\{ \uparrow,\downarrow \}$ denotes the spin degree of freedom and 
noting that $\mathbf{L} \cdotp \mathbf{S}=L_z S_z + L_{+} S_{-} + L_{-} S_{+}$, where $L_{\pm}= L_{x} \pm i L_{y}$ and 
$S_{\pm}=\frac{1}{2}(S_{x}\pm i S_{y})$, it is straightforward to calculate the matrix elements 
$(H_{so}^{ABC})_{\mu,\nu} = \langle \Psi_{\mu} |\hat{H}_{SO}^{atomic} |\Psi_{\nu}\rangle$ in the symmetry basis introduced earlier. 
Using the notation $S_{\pm}^{\tau}=\frac{1}{2}(S_{x}\pm i \tau \cdotp S_{y})$,  where $\tau=+1\, (-1)$ corresponds to the 
$K$ ($K'$) point of the BZ, the result is shown in Table \ref{intrinsic-SO-trilayer}. 
\begin{table}[ht]
\begin{tabular}{l|cccccc}\hline\hline\vspace*{-0.8em}
  & & & & & &\\
%\vspace*{-0.8em}
SOC  &         $\Psi_{\Gamma_{A_1}}$          &          $\Psi_{\Gamma_{A_2}}$          &          $\Psi_{\Gamma_{E_{1,1}}}$          &
          $\Psi_{\Gamma_{E_{1,2}}}$          &          $\Psi_{\Gamma_{E_{2,1}}}$          &          $\Psi_{\Gamma_{E_{2,2}}}$       \\ 
\vspace*{-0.8em}          
 & & & & & &\\           
          \hline\vspace*{-0.8em}
  & & & & & &\\
\vspace*{-0.8em}      
$\Psi_{\Gamma_{A_1}}$   &            $0$              &        $\lambda_{1/2} S_{z}^{\tau}$        &        $\lambda_{1/3}^{E_1} S_{+}^{\tau}$        &
        $\lambda_{1/3}^{E_1} S_{-}^{\tau}$        &        $-\lambda_{1/3}^{E_2} S_{+}^{\tau}$        &        $\lambda_{1/3}^{E_2} S_{-}^{\tau}$        \\ 
  & & & & & &\\
\vspace*{-0.8em}
$\Psi_{\Gamma_{A_2}}$   &        $\lambda_{1/2} S_{z}$        &        $0$        &        $-\lambda_{2/3}^{E_1} S_{+}^{\tau}$        &
        $\lambda_{2/3}^{E_1} S_{-}^{\tau}$        &        $\lambda_{2/3}^{E_2} S_{+}^{\tau}$        &         $\lambda_{2/3}^{E_2} S_{-}^{\tau}$        \\ 
  & & & & & &\\      
\vspace*{-0.8em}      
$\Psi_{\Gamma_{E_{1,1}}}$    &        $\lambda_{1/3}^{E_{1}} S_{-}^{\tau}$        &        $-\lambda_{2/3}^{E_1} S_{-}^{\tau}$        &           $0$           &
            $0$             &        $\lambda_{3/3}^{z} S_{z}$        &        $\lambda_{3/3} S_{+}^{\tau}$        \\
  & & & & & &\\          
\vspace*{-0.8em}
$\Psi_{\Gamma_{E_{1,2}}}$    &        $\lambda_{1/3}^{E_1} S_{+}^{\tau}$        &          $\lambda_{2/3}^{E_1} S_{+}^{\tau}$        &            $0$           &
            $0$             &           $\lambda_{3/3} S_{-}^{\tau}$        &        $\lambda_{3/3}^{z} S_{z}$      \\
  & & & & & &\\            
\vspace*{-0.8em}
$\Psi_{\Gamma_{E_{2,1}}}$    &        $-\lambda_{1/3}^{E_2} S_{-}^{\tau}$        &         $\lambda_{2/3}^{E_2} S_{-}^{\tau}$         &
        $\lambda_{3/3}^{z}\, S_{z}$        &        $\lambda_{3/3} S_{+}^{\tau}$        &              $0$             &             $0$             \\
  & & & & & &\\      
\vspace*{-0.8em}
$\Psi_{\Gamma_{E_{2,2}}}$    &        $\lambda_{1/3}^{E_2} S_{+}^{\tau}$        &         $\lambda_{2/3}^{E_2} S_{+}^{\tau}$         &
        $\lambda_{3/3} S_{-}^{\tau}$        &        $\lambda_{3/3}^{z} S_{z}$       &             $0$              &             $0$               \\       
  & & & & & &      \\
\hline\hline
\end{tabular}
\caption{Intrinsic spin-orbit Hamiltonian $\tau\, H_{so}^{ABC}$  in the symmetry basis. Here $\tau=+1 (-1)$ corresponds to the $K$ 
($K'$) point.}
\label{intrinsic-SO-trilayer}
\end{table}

The SOC Hamiltonian shown in Table \ref{intrinsic-SO-trilayer} is the main result of this section. 
Explicit expressions in terms of SK parameters for the coupling constants appearing in Table 
\ref{intrinsic-SO-trilayer} can be found in Table \ref{tbl:SlatKost-SOC}.  
In contrast to previous works where SOC in ABC trilayer was discussed\cite{konschuh-phd,guinea-so}, 
we find that the number of   SOC parameters %in ABC trilayer graphene 
is seven\cite{two-soc-param}. 
As we will show, $\lambda_{1/3}^{E_{1}},\lambda_{1/3}^{E_{2}},\lambda_{2/3}^{E_{1}}$ and $\lambda_{2/3}^{E_{2}}$ 
are related to  interlayer SOC and calculations  which are based on the symmetry properties 
of low-energy effective Hamiltonians may not capture them. 
The $\lambda_{3/3}$ parameter ensures that the otherwise fourfold degeneracy of the split-off states at the $K$ 
point is lifted, as it is dictated by  general group-theoretical considerations\cite{dresselhaus-paper,dresselhaus-book}.
These five parameters are proportional to the product $V_{p d \pi} V_{p d \sigma}$ and  they could not be obtained
considering only the $d_{xz}$, $d_{yz}$ orbitals and in-plane SOC. The remaining two SOC parameters, $\lambda_{1/2}$ and 
$\lambda_{3/3}^{z}$ are proportional to $V_{p d \pi}^2$ and describe in-plane SOC. 

\begin{table}[htb]
\begin{tabular}{l|cccccc}\hline\hline\vspace{-0.8em}
 & & & & & &\\
SOC  &         $\Psi^{A1}_{\rm eff}$          &          $\Psi^{B3}_{\rm eff}$          &          $\Psi^{B1}_{\rm eff}$          &
          $\Psi^{A2}_{\rm eff}$          &          $\Psi^{B2}_{\rm eff}$          &          $\Psi^{A3}_{\rm eff}$          \\ 
\vspace*{-0.8em}          
 & & & & & &\\
          \hline \vspace*{-0.8em} 
 & & & & & &\\
\vspace*{-0.8em}      
$\Psi^{A1}_{\rm eff}$   &         ${\lambda}_{1/2}\, S_{z}$          &        $0$        &        ${\lambda}_{2}\, S_{+}^{\tau}$        &
        $\lambda_{1}\, S_{+}^{\tau}$        &        $\lambda_{4}\, S_{-}^{\tau}$        &        $\lambda_{3}\, S_{-}^{\tau}$        \\ 
 & & & & & &\\        
\vspace*{-0.8em}
$\Psi^{B3}_{\rm eff}$   &        $0$        &       $-\lambda_{1/2}\, S_{z}$      &        $-\lambda_{3}\, S_{+}^{\tau}$        &
        $-\lambda_{4}\, S_{+}^{\tau}$        &        $-\lambda_{1}\, S_{-}^{\tau}$        &         $-\lambda_{2}\, S_{-}^{\tau}$        \\ 
 & & & & & &\\        
\vspace*{-0.8em}      
$\Psi^{B1}_{\rm eff}$    &        ${\lambda}_{2}\, S_{-}^{\tau}$        &        $-\lambda_{3}\, S_{-}^{\tau}$        &            $-{\lambda}_{3/3}^{z}\, S_{z}$           &
           $0$            &        $-\lambda_{3/3}\, S_{+}^{\tau}$        &        $0$        \\
 & & & & & &\\           
\vspace*{-0.8em}
$\Psi^{A2}_{\rm eff}$    &        $\lambda_{1}\, S_{-}^{\tau}$        &          $-\lambda_{4}\, S_{-}^{\tau}$        &             $0$           &
           $\lambda_{3/3}^{z}\, S_{z}$              &           $0$        &        $\lambda_{3/3}\, S_{+}^{\tau}$      \\
 & & & & & &\\           
\vspace*{-0.8em}
$\Psi^{B2}_{\rm eff}$    &        $\lambda_{4}\, S_{+}^{\tau}$        &         $-\lambda_{1}\, S_{+}^{\tau}$         &        $-\lambda_{3/3}\, S_{-}^{\tau}$        &
        $0$        &           $-\lambda_{3/3}^{z}\, S_{z}$            &               $0$             \\
 & & & & & &\\        
\vspace*{-0.8em}
$\Psi^{A3}_{\rm eff}$    &        $\lambda_{3}\, S_{+}^{\tau}$        &         $-\lambda_{2}\, S_{+}^{\tau}$         &        $0$        &
        $\lambda_{3/3}\, S_{-}^{\tau}$       &             $0$              &        $\lambda_{3/3}^{z}\, S_{z}$          \\       
 & & & & & &\\        
\hline\hline
\end{tabular}
\caption{$ \tau\, H_{so}^{ABC}$ in the basis of effective $p_z$ orbitals. 
}
\label{intrinsic-SO-pzeff}
\end{table}

\begin{table*}[htb]
\begin{tabular}{l|l}\hline\hline\vspace*{-0.8em}
 & \\
$|\Psi_{\Gamma_{A_{1}}}\rangle=|\overline{\Psi_{1,0,0}^{1,3}}\rangle+
              \frac{3}{\sqrt{2}}\frac{V_{p d \pi}}{\delta\vareps_{pd}-\gamma_2} |\Psi_{2,1,-1}^{3,1}\rangle 
           + \frac{\tilde{V}_{p d \sigma}}{\delta\vareps_{pd}-\tilde{\gamma}_2} |\Psi_{2,0,0}^{1,3}\rangle$           &
\,\, $\lambda_{1/2}=-\frac{9}{2} \,\xi_d\, \frac{(V_{p d\pi})^2}{(\delta\vareps_{pd}+\gamma_2)(\delta\vareps_{pd}-\gamma_2)}$ \\
\vspace*{-0.8em}
 & \\
\vspace*{-0.8em}
$|\Psi_{\Gamma_{A_{2}}}\rangle=|\Psi_{1,0,0}^{1,3}\rangle-
           \frac{3}{\sqrt{2}}\frac{V_{p d \pi}}{\delta\vareps_{pd}+\gamma_2} |\overline{\Psi_{2,1,-1}^{3,1}}\rangle 
           - \frac{\tilde{V}_{p d \sigma}}{\delta\vareps_{pd}+\tilde{\gamma}_2} |\overline{\Psi_{2,0,0}^{1,3}}\rangle$    &   
\,\, $\lambda_{1/3}^{E_1}=\frac{3 \sqrt{3}}{2}\, \xi_d\, \frac{V_{p d \pi}}{(\delta\vareps_{pd}+\gamma_1)} 
             \left(\frac{\tilde{V}_{p d \sigma}}{\delta\vareps_{pd}-\tilde{\gamma}_2}-
              \frac{V_{p d \sigma}}{\delta\vareps_{pd}-\gamma_2}\right)$ \\         
 & \\              
\vspace*{-0.8em}
$|\Psi_{\Gamma_{E_{1,1}}}\rangle=|\Psi_{1,0,0}^{2,1}\rangle+\frac{1}{\delta\vareps_{pd}+\gamma_1}
           \left[\frac{V_{p d \sigma}}{\sqrt{2}}|\overline{\Psi_{2,0,0}^{2,1}}\rangle 
           -\frac{3 V_{p d \pi}}{2}  |\overline{\Psi_{2,1,-1}^{1,2}}\rangle \right]$    &   
\,\, $\lambda_{1/3}^{E_2}=\frac{3 \sqrt{3}}{2} \,\xi_d \,\frac{V_{p d \pi}}{(\delta\vareps_{pd}-\gamma_1)}
                   \left(\frac{\tilde{V}_{p d \sigma}}{\delta\vareps_{pd}-
                   \tilde{\gamma}_2}+\frac{V_{p d \sigma}}{\delta\vareps_{pd}-\gamma_2}\right)$ \\
 & \\                   
\vspace*{-0.8em}
$|\Psi_{\Gamma_{E_{1,2}}}\rangle=|\Psi_{1,0,0}^{3,2}\rangle+\frac{1}{\delta\vareps_{pd}+\gamma_1}
           \left[\frac{V_{p d \sigma}}{\sqrt{2}} |\overline{\Psi_{2,0,0}^{3,2}}\rangle 
           -\frac{3 V_{p d \pi}}{2}  |\overline{\Psi_{2,1,-1}^{2,3}}\rangle\right]$     &  
\,\, $\lambda_{2/3}^{E_1}=\frac{3 \sqrt{3}}{2} \,\xi_d \,\frac{V_{p d \pi}}{(\delta\vareps_{pd}+\gamma_1)}
                      \left(\frac{\tilde{V}_{p d \sigma}}{\delta\vareps_{pd}+\tilde{\gamma}_2}+
                      \frac{V_{p d \sigma}}{\delta\vareps_{pd}+\gamma_2}\right)$  \\
 & \\                      
\vspace*{-0.8em}
$|\Psi_{\Gamma_{E_{2,1}}}\rangle=|\overline{\Psi_{1,0,0}^{2,1}}\rangle+\frac{1}{\delta\vareps_{pd}-\gamma_1}
            \left[\frac{3 V_{p d \pi}}{2} |\Psi_{2,1,-1}^{1,2}\rangle-
            \frac{V_{p d \sigma}}{\sqrt{2}}|\Psi_{2,0,0}^{2,1}\rangle\right]$   &  
\,\, $\lambda_{2/3}^{E_2}=\frac{3 \sqrt{3}}{2} \,\xi_d\, \frac{V_{p d \pi}}{(\delta\vareps_{pd}-\gamma_1)}
                    \left(\frac{\tilde{V}_{p d \sigma}}{\delta\vareps_{pd}+\tilde{\gamma}_2}-
                    \frac{V_{p d \sigma}}{\delta\vareps_{pd}+\gamma_2}\right)$ \\
 & \\                    
\vspace*{-0.8em}
$|\Psi_{\Gamma_{E_{2,2}}}\rangle=|\overline{\Psi_{1,0,0}^{3,2}}\rangle+\frac{1}{\delta\vareps_{pd}-\gamma_1}
            \left[\frac{3 V_{p d \pi}}{2} |\Psi_{2,1,-1}^{2,3}\rangle-
            \frac{V_{p d \sigma}}{\sqrt{2}}|\Psi_{2,0,0}^{3,2}\rangle\right]$   &  
\,\, $\lambda_{3/3}={3 \sqrt{3}} \,\xi_d\,
              \frac{V_{p d \pi} V_{p d \sigma}}{(\delta\vareps_{pd}+\gamma_1)(\delta\vareps_{pd}-\gamma_1)}$  \\
& \\              
\vspace*{-0.8em}   
    &\,\,  $\lambda_{3/3}^{z}=-\frac{9}{2}\, \xi_d \, 
           \frac{(V_{p d \pi})^2}{(\delta\vareps_{pd}+\gamma_1)(\delta\vareps_{pd}- \gamma_1)}$ \\
& \\
\hline\hline
\end{tabular}
\caption{Symmetry basis functions (left column) and intrinsic spin-orbit matrix elements (right column) in terms of  
Slater-Koster parameters. Here $\xi_d$ is the angular momentum resolved atomic SOC strength,
$V_{d d \sigma}= V_{d d \sigma}^{A1,B3}$, $\tilde{V}_{p d \sigma}=V_{p d \sigma}^{A1,B3}$, 
$V_{p d \sigma}=V_{p d \sigma}^{A2,B1}=V_{p d \sigma}^{A3,B2}$, 
$\delta\vareps_{pd}=\vareps_p-\vareps_d$, $\tilde{\gamma}_2=\gamma_2+V_{d d \sigma}$ 
and we assumed that $V_{p d \pi}=V_{p d \pi}^{Ai,Bi}$, $i=1,2,3$. Although the basis functions shown in the right 
are not normalized, the SOC parameters are correct in the lowest order of the products 
of the small parameters 
${\tilde{V}_{p d \sigma}} / {(\delta\vareps_{pd}\pm\tilde{\gamma}_2)}$, 
${{V}_{p d \sigma}} / {(\delta\vareps_{pd}\pm{\gamma}_2)}$, 
${V_{p d \pi}} / {(\delta\vareps_{pd}\pm\gamma_1)}$
}
\label{tbl:SlatKost-SOC}
\end{table*}

The physical meaning of these SOC parameters is probably more transparent if one rotates  $H_{so}^{ABC}$ 
into the basis of on-site effective $p_z$ orbitals (see Appendix \ref{sec:on-site-transf}) . These basis vectors result from  
the admixture of an  on-site $p_z$ orbital $|\Psi_{1,0}^{\alpha j}\rangle$ with large weight  and  
$|\Psi_{2,\pm 1}^{\alpha' j}\rangle$,  $|\Psi_{2, 0}^{\alpha j}\rangle$ ($\alpha\neq\alpha'$)   with small weight.  
$H_{so}^{ABC}$ in the on-site effective $p_z$  basis is shown in Table \ref{intrinsic-SO-pzeff}. 
The SOC parameters $\lambda_1,\lambda_2,\lambda_3$ and $\lambda_4$ are linear combinations of 
$\lambda_{1/3}^{E_{1}}$, $\lambda_{1/3}^{E_{2}}$, $\lambda_{2/3}^{E_{1}}$ and $\lambda_{2/3}^{E_{2}}$, 
the explicit relations are given in  Section \ref{sec:SO-ABC-SL}. In Section \ref{sec:SO-ABC-SL} we will also 
make  predictions which might be useful to guide the fitting procedure if results of DFT calculations are fitted with a 
TB model, as  e.g. in Ref.~\onlinecite{jfabian-bilayer}. 

Since the lattice of Bernard stacked bilayer graphene has the same symmetry group as ABC trilayer, 
the considerations made in this section can be easily applied to  bilayer graphene as well. 
A brief summary of the  bilayer calculations is given in Section \ref{sec:bilayer}. 
The importance of the bilayer results is that they can be compared with 
the numerical calculations of Ref.~\onlinecite{jfabian-bilayer}. 
Based on this comparison   we will be able to   
estimate the values of five of the seven SOC parameters of  ABC trilayer.

\section{Intrinsic SOC in bilayer graphene}
\label{sec:bilayer}

In this Section we give a brief summary of our TB calculations for the intrinsic SOC in bilayer 
graphene and compare the results to the DFT computations of Ref.~\onlinecite{jfabian-bilayer}. 
The low-energy states of bilayer are also found at the $K$ and $K'$ points of the BZ, hence  
the calculation follows the same steps as in Section \ref{sec:intrinsic-soc}: (i) first  
we obtain  the basis states of the symmetry basis  
$
\{ |\Psi_{\Gamma_{A_{1}}}\rangle, |\Psi_{\Gamma_{A_{2}}}\rangle, |\Psi_{\Gamma_{E_{1,1}}}\rangle,
|\Psi_{\Gamma_{E_{1,2}}}^{}\rangle \},
$
and (ii)  we calculate the matrix elements of $\hat{H}_{SO}^{atomic}$. 
Note that in the case of bilayer graphene there is only one pair of bands which transforms as 
the two-dimensional representation $\Gamma_{E}$.

For easier comparison we adopt the notation of Ref.~\onlinecite{jfabian-bilayer} for the SOC parameters.  
As it is shown in Table \ref{tbl:intrinsic-SO-bilayer}, the SO Hamiltonian in the basis of the effective 
$p_z$ orbitals can be written in a form  which, apart from a unitary transformation,
agrees with the result given in Table IV of Ref.~\onlinecite{jfabian-bilayer} (see Appendix \ref{sec:irrep-basis-funct} for 
details).
\begin{table}[htb]
\begin{tabular}{l|cccc}\hline\hline\vspace*{-0.8em}
 & & & &\\
SOC  &         $\Psi^{A1}_{\rm eff}$          &          $\Psi^{B1}_{\rm eff}$          &          $\Psi^{A2}_{\rm eff}$          &
          $\Psi^{B2}_{\rm eff}$                    \\ 
 \vspace*{-0.8em}
 & & & &\\
 \hline \vspace*{-0.8em}
 & & & &\\
 \vspace*{-0.8em}
$\Psi^{A1}_{\rm eff}$   &         ${\lambda}_{\rm I 2}\, S_{z}$          &        $\lambda_0\, S_{-}^{\tau}$        &        ${\lambda}_{4}^{bi}\, S_{+}^{\tau}$        &
        $0$          \\ 
 & & & &\\       
\vspace*{-0.8em}
$\Psi^{B1}_{\rm eff}$   &        $\lambda_0\, S_{+}^{\tau}$        &       $-\lambda_{\rm I 1}\, S_{z}$      &        $0$        &
        $-\lambda_{4}^{bi}\, S_{+}^{\tau}$               \\ 
 & & & &\\         
\vspace*{-0.8em}      
$\Psi^{A2}_{\rm eff}$    &        ${\lambda}_{4}^{bi}\, S_{-}^{\tau}$        &        $0$        &            ${\lambda}_{\rm I 1}\, S_{z}$           &
           $-\lambda_0\, S_{-}^{\tau}$                \\
 & & & &\\           
\vspace*{-0.8em}
$\Psi^{B2}_{\rm eff}$    &        $0$        &          $-\lambda_{4}^{bi}\, S_{-}^{\tau}$        &           $-\lambda_0\, S_{+}^{\tau}$           &
           $-\lambda_{\rm I 2}\, S_{z}$                  \\   
 & & & &\\           
\hline\hline
\end{tabular}
\caption{Intrinsic spin-orbit Hamiltonian $\tau \, H_{so}^{AB}$ of bilayer graphene in the basis of effective $p_z$ orbitals.  
}
\label{tbl:intrinsic-SO-bilayer}
\end{table}
In terms of the SK hoppings, the SOC parameters read:
\begin{eqnarray}
\lambda_0 &= & 3 \sqrt{3}\,\xi_d\, \frac{V_{p d \pi} V_{p d \sigma}}{\delta\vareps_{pd}^2-\tilde{\gamma}_1^2}\frac{\tilde{\gamma}_1}{\vareps_{pd}}; 
\hspace{2em} 
\lambda_{\rm I 1} = \frac{9}{2}\,\xi_d\,\frac{V_{p d \pi}^2}{\delta\vareps_{pd}^2};\nonumber\\
\lambda_4^{bi} &=& 3 \sqrt{3}\,\xi_d\, \frac{V_{p d \pi} V_{p d \sigma}}{\delta\vareps_{pd}^2-\tilde{\gamma}_1^2}; 
\hspace{2em}
\lambda_{\rm I 2} = \frac{9}{2}\,\xi_d\, \frac{V_{p d \pi}^2}{\delta\vareps_{pd}^2-\gamma_1^2} 
\label{bilayer-SOC-SL}
\end{eqnarray}
where $\tilde{\gamma}_1=\gamma_1+V_{d d \sigma}$.  Looking at the expressions given in (\ref{bilayer-SOC-SL}), one can 
make the following observations: 
(i) Since $\delta\vareps_{pd} = \vareps_p-\vareps_d < 0$ and $\tilde{\gamma}_1 > 0$, 
one can expect that the sign of $\lambda_0$ and $\lambda_4^{bi}$ will be different; 
(ii) $|\lambda_0| < |\lambda_4^{bi}|$ because $\frac{\tilde{\gamma}_1}{|\vareps_{pd}|} < 1$;
(iii) $\lambda_{\rm I 1}$ and $\lambda_{\rm I 2}$ have the same sign and they are approximately of the same
magnitude because $\gamma_1^2/\delta\vareps_{pd}^2\ll 1$.

By fitting the band structure obtained from DFT calculations 
with their tight binding model, the authors of Ref.~\onlinecite{jfabian-bilayer} 
found the following values for the bilayer SOC parameters: $2 \lambda_{\rm I 1} = 24 \mu{\rm eV}$, 
$2 \lambda_{\rm I 2} = 20 \mu{\rm eV}$, $\lambda_0 = 5 \mu{\rm eV}$, $\lambda_4 = -12 \mu{\rm eV}$.
These numbers are in qualitative agreement with the predictions 
%for the magnitude and mutual relations of the SOC parameters 
that we made below Eq.~(\ref{bilayer-SOC-SL}) for the SOC parameters.
If, in addition, one assumes that $V_{p d \pi} V_{p d \sigma} < 0$ then according to  (\ref{bilayer-SOC-SL}) 
three of the parameters ($ \lambda_{\rm I 1}$, $ \lambda_{\rm I 2}$, $ \lambda_{0}$) 
should have the same sign, which would again agree with the 
results of Ref.~\onlinecite{jfabian-bilayer}.

\section{SOC parameters for trilayer graphene in terms of SK hoppings}
\label{sec:SO-ABC-SL}

We are now ready to make predictions for five  of the seven ABC trilayer SOC parameters. 
To this end, we first express the SOC parameters in the effective $p_z$ orbital basis in terms of the 
SOC parameters obtained in the symmetry basis. 
Moreover, using the formulae given in Table \ref{tbl:SlatKost-SOC} 
for  $\lambda_{1/3}^{E_{1}}$, $\lambda_{1/3}^{E_{2}}$, $\lambda_{2/3}^{E_{1}}$ and  
$\lambda_{2/3}^{E_{2}}$, 
one can also express $\lambda_1$, $\lambda_2$, $\lambda_3$ and $\lambda_4$ in terms of  the SK hoppings. 
Taking into account the results of the previous section, this then facilitates  making predictions for
the numerical values of the ABC trilayer SOC parameters. 

First, the SOC parameters in terms of the SK hoppings:
\begin{subequations}
\begin{eqnarray}
\lambda_1 &=&(\lambda_{1/3}^{E_1}+\lambda_{2/3}^{E_2}-\lambda_{1/3}^{E_2}-\lambda_{2/3}^{E_1})/2\nonumber\\ 
           &\approx& \eta\left(-V_{p d \sigma}+\tilde{V}_{p d\sigma}\frac{\gamma_2}{\delta\vareps_{pd}}\right),\label{lambda1}\\
\lambda_2 &=& (\lambda_{1/3}^{E_1}+\lambda_{1/3}^{E_2}-\lambda_{2/3}^{E_1}-\lambda_{2/3}^{E_2})/2 \nonumber\\
          &\approx& \eta \left(V_{p d \sigma}\frac{\gamma_1}{\delta\vareps_{pd}}+\tilde{V}_{p d\sigma}\frac{\gamma_2}{\delta\vareps_{pd}}\right),\\
\lambda_3 &=& (\lambda_{1/3}^{E_1}+\lambda_{1/3}^{E_2}+\lambda_{2/3}^{E_1}+\lambda_{2/3}^{E_2})/2 \nonumber\\
          &\approx& \eta\left(V_{p d \sigma}\frac{\gamma_1}{\delta\vareps_{pd}}\frac{\gamma_2}{\delta\vareps_{pd}}+\tilde{V}_{p d\sigma}\right),\\
\lambda_4 &=& (\lambda_{1/3}^{E_1}+\lambda_{1/3}^{E_2}+\lambda_{2/3}^{E_1}+\lambda_{2/3}^{E_2})/2 \nonumber\\
          &\approx& \eta \left(V_{p d \sigma}\frac{\gamma_2}{\delta\vareps_{pd}}+\tilde{V}_{p d\sigma}\frac{\gamma_1}{\delta\vareps_{pd}}\right),
\end{eqnarray}
\label{pzeff-SOC-SL}
\end{subequations}
where $\eta=3 \sqrt{3}\,\xi_d\,\frac{V_{pd\pi}}{\delta\vareps_{pd}^2-\gamma_1^2}$. 
% We remind the reader that 
% $V_{p d \sigma}=V_{p d \sigma}^{A2,B1}=V_{p d \sigma}^{A3,B2}$ and $\tilde{V}_{p d \sigma}=V_{p d \sigma}^{A1,B3}$ 
% (Table \ref{tbl:SlatKost-SOC}).
Similarly to the bilayer case,  looking at Table \ref{tbl:SlatKost-SOC} and Eqs.~(\ref{pzeff-SOC-SL}) one can make 
the following observations:
(i)  One would expect that  $\lambda_{1/2}\approx \lambda_{3/3}^z$,
(ii) $|\lambda_{3/3}|\approx|\lambda_1|$  and  assuming that 
$V_{p d \sigma}/\tilde{V}_{p d \sigma} \propto \gamma_1/\gamma_2$  one finds that 
$|\lambda_1|>|\lambda_2|>|\lambda_3|>|\lambda_4|$, and (iii) $\lambda_4$ has opposite sign 
from  $\lambda_3$ and similarly for $\lambda_{3/3}$ and $\lambda_{1}$ because the second
term in the expression for $\lambda_{1}$ in Eq.~(\ref{lambda1}) can be neglected with respect
to the first one.

Comparing the expressions in terms of the SK hoppings given in Eqs.~(\ref{pzeff-SOC-SL}) with the corresponding 
ones for the bilayer case in (\ref{bilayer-SOC-SL}), the following estimates can be made: 
$2 \lambda_{1/2}^{z}\approx 2 \lambda_{3/3}^z\approx 20 \mu{\rm eV}$, 
$|\lambda_{3/3}| \approx |\lambda_1| \approx 10 \mu{\rm eV}$ and $ |\lambda_2| \approx 5 \mu{\rm eV}$.
Since $\lambda_3$ and $\lambda_4$ are proportional to $\tilde{V}_{p d \sigma}$  
(assuming $V_{p d \sigma}/\tilde{V}_{p d \sigma} \propto \gamma_1/\gamma_2$) 
which is unknown, we cannot  give  a numerical estimate for their value. 
One would expect that they are much smaller than $\lambda_1$ and $\lambda_2$ because 
$\tilde{V}_{p d \sigma}$ corresponds to a remote, and presumably weak $p-d$ hopping between the $A1$ and $B3$ sites.

\section{Effective SOC Hamiltonian}
\label{sec:eff_so_ham}

The calculations in the previous sections are valid, strictly speaking, only at the $K$ point of the Brillouin zone. 
To obtain the Hamiltonian in the vicinity of the $K$ point, where the states close to the Fermi energy can be found, 
one can perform a $\mathbf{k}\cdotp \mathbf{p}$ expansion of the bands. 
We neglect the weak $\mathbf{k}$ dependence of the SOC \cite{jfabian-bilayer}, 
hence the total Hamiltonian of the system can be written as 
$\hat{H}_{ABC}=\hat{H}_{\mathbf{k}\cdotp\mathbf{p}}^{ABC} +\tau\, \hat{H}_{so}^{ABC}$.  
Here ${\hat H}_{\mathbf{k}\cdotp\mathbf{p}}^{ABC}$ is  the $\mathbf{k}\cdotp\mathbf{p}$ Hamiltonian obtained 
without taking into account the SOC,   whereas ${\hat H}_{so}^{ABC}$ is the SOC Hamiltonian calculated at the $K$ point. 
${\hat H}_{\mathbf{k}\cdotp\mathbf{p}}^{ABC}$ 
has been published  before, see e.g. Refs.~\onlinecite{mccann-abctrilayer,macdonald-1}.

To study the low energy physics however, in which we are 
primarily interested, the use of ${\hat H}_{\mathbf{k}\cdotp\mathbf{p}}^{ABC}$ is not convenient, 
since it includes four bands that are split-off from the Fermi energy of the (undoped) ABC trilayer 
by the large energy scale $\approx \pm \gamma_1$\cite{mccann-abctrilayer,macdonald-1}. 
Therefore we derive an effective two component (or, including the spin, four component) 
Hamiltonian $\hat{H}^{\rm eff}_{ABC}$ which describes the hopping between atomic sites 
$A1$ and $B3$. To this end we again employ  the Schrieffer-Wolff transformation 
and keep all  terms which are third order or
less in the momentum $\pi$, $\pi^{\dagger}$ and first order in the SOC constants. 
Here $\pi=-(i p_x+ \tau p_y)$, where $\tau=1 (-1)$ for valley $K$ ($K'$)  
and the momenta $p_x,p_y$ are measured from the $K$ ($K'$)  point of the BZ, 
see Fig.~\ref{fig1}(c). Keeping terms up to 
third order in $\pi$, $\pi^{\dagger}$ is essential to reproduce the important 
features of the low-energy band structure\cite{mccann-abctrilayer,macdonald-1}, 
such as the band degeneracy and the trigonal warping. The necessary formulae for 
the matrix elements of the effective Hamiltonian 
can be found in Ref.~\onlinecite{winkler-book}. We treat $\gamma_1$ as a large energy scale
with respect to $v_0 \pi $, $|\gamma_3|$, $\gamma_4$, $\gamma_2$ and the typical energies $E$ 
we are interested in and keep only the leading order for the terms involving $\gamma_2$, $v_3$ and $v_4$. 
(The velocities $v_i$ are given by  $v_i=(\sqrt{3}/2) a \gamma_i/\hbar$, where $a=0.246\,{\rm nm}$ is the 
lattice constant of graphene.)
For the folding down  of the 
full Hamiltonian we use the form of $\hat{H}_{ABC}$ in the symmetry basis because in this case
all the large matrix elements are on the diagonal 
and therefore the quasidegenerate perturbation approach is expected to work well. 
Once we obtain the effective Hamiltonian $\hat{H}^{\rm eff}_{ABC}$ in the symmetry basis we rotate 
it into the  basis of   effective $p_z$ orbitals centered on atomic sites $A1$ and $B3$ 
because  $\hat{H}^{\rm eff}$ assumes a simpler form in this basis. 
Explicitly, one can write $\hat{H}^{\rm eff}_{ABC} = \hat{H}^{\rm eff}_{el} + \hat{H}^{\rm eff}_{so}$, 
where the electronic part is given by 
\begin{subequations}
\begin{eqnarray}
{\hat{H}}^{\rm \, eff }_{el} &=& {\hat{H}}_{chir} + {\hat{H}}_{3w} + \hat{H}_{\gamma_2} + {\hat{H}}_{v_4},\nonumber\\
{\hat{H}}_{chir} &=& \frac{v_0^3}{\gamma_1^2}\left(
\begin{array}{cc}
0 & \left( {\pi }^{\dag }\right)^{3} \\
{\pi^{3}} & 0
\end{array} \right),\\
{\hat{H}}_{3w} &=&
-\frac{v_0 v_3}{\gamma_1} [\pi^{\dagger}\pi + \pi\pi^{\dagger}]
 \left(
\begin{array}{cc}
0 & 1 \\
1 & 0
\end{array}
\right),\\
\hat{H}_{\gamma_2} &=& \gamma_2 \left(1 -\frac{1}{2}\frac{v_0^2}{\gamma_1^2}[\pi^{\dagger}\pi + \pi\pi^{\dagger}]\right)
\left(
\begin{array}{cc}
 0 & 1 \\
 1 & 0 
\end{array}
\right),\\
{\hat{H}}_{v_4} &=&
-\frac{2 v_0 v_4}{\gamma_1}
\left(
\begin{array}{cc}
\pi^{\dagger} \pi & 0 \\
0 & \pi \pi^{\dagger}
\end{array}
\right).
\end{eqnarray}
\label{inv_sym_el_eff}
\end{subequations}
We note that when applying the Schrieffer-Wolff transformation we did not assume that $\pi$ and $\pi^{\dagger}$ commute, 
therefore the Hamiltonians (\ref{inv_sym_el_eff}) and (\ref{inv_sym_intr_so_eff})  
are valid \emph{in the presence of finite external magnetic field} as well.  
In zero magnetic field, using the notation $p^2=\pi\pi^{\dagger}=\pi^{\dagger}\pi$,   
the Hamiltonian (\ref{inv_sym_el_eff})  simplifies to the corresponding results  in 
Refs.~\onlinecite{mccann-abctrilayer,macdonald-1}. 

The effective SOC Hamiltonian is
\begin{subequations}
\begin{eqnarray}
 {\hat{H}}^{\rm  \,eff }_{so} &=&  {\hat{H}}_{so}^{mn} + {\hat{H}}_{so}^{(1)} + {\hat{H}}_{so}^{(2)} + {\hat{H}}_{so}^{(3)},\nonumber\\
{\hat{H}}_{so}^{mn} &=&  \tau\,\lambda_{1/2} S_{z} \sigma_z,\\ 
 {\hat{H}}_{so}^{(1)} &=&  -\tau\,\tilde{\lambda}_1\, \frac{v_0}{\gamma_1}\, [S_{-}^{\tau} \pi^{\dagger}  + S_{+}^{\tau} \pi ] \sigma_{z},\\  
{\hat{H}}_{so}^{(2)} &=&  \tau\,\frac{v_0 v_3}{\gamma_1^2} \lambda_{3/3} [ S_{+}^{\tau} (\pi^{\dagger})^2  + S_{-}^{\tau} \pi^2] \sigma_{z},\\ 
{\hat{H}}_{so}^{(3)} &=&
- \tau\,\frac{v_0^2}{\gamma_1^2} (\lambda_{3/3}^{z}-\lambda_{1/2}) S_z 
\left(
\begin{array}{cc}
\pi^{\dagger} \pi & 0 \\
0 & -\pi \pi^{\dagger}
\end{array}
\right).
\label{inv_sym_intr_so_eff}
\end{eqnarray}
\label{inv_sym_intr_so_eff}
\end{subequations}
Here the Pauli matrix $\sigma_z$ acts in the space of  $\{ A1, B3 \}$  sites and 
$\tilde{\lambda}_1=\lambda_1+\lambda_3 (v_3/v_0) + \lambda_2 (v_4/v_0) \approx \lambda_1$. 
At low energies $v_0 ^2 p^2, v_0 v_3 p^2 \ll \gamma_1^2$ and the corresponding terms in (\ref{inv_sym_intr_so_eff}) can be neglected. 
The first term, ${\hat{H}}_{so}^{mn}$ is the well known SO Hamiltonian of monolayer graphene\cite{kane-mele} 
and describes the leading contribution to SOC. The next term, ${\hat{H}}_{so}^{(1)}$ is the most important  momentum dependent 
contribution close to the $K$ point. 
Keeping only ${\hat{H}}_{so}^{(mn)}$ and ${\hat{H}}_{so}^{(1)}$ the effective SOC Hamiltonian can be written in a more compact form as 
\begin{equation}
  {\hat{H}}^{\rm  \,eff }_{so}=\tau\, \left[\lambda_{1/2} S_z + 
 \tilde{\lambda}_1\frac{v_0}{\gamma_1}\left(\mathbf{S}\times\mathbf{p}\right)_z  \right] \sigma_z,
 \label{so-ham-compact}
\end{equation}
where $\mathbf{p}=(p_x,p_y)$. 
We note that Eq.~(\ref{so-ham-compact}) also describes the effective SOC Hamiltonian of bilayer graphene with $\lambda_{1/2}$ 
replaced by $-\lambda_{\rm I 1}$ and 
$\tilde{\lambda}_1=-\lambda_4^{bi}+\lambda_0 (v_4/v_0) \approx -\lambda_4^{bi}$ 
(for  $\lambda_{\rm I 1}$, $\lambda_0$ and $\lambda_4^{bi}$ see Sect.~\ref{sec:bilayer}).

\begin{figure}[ht]
\includegraphics[scale=0.45]{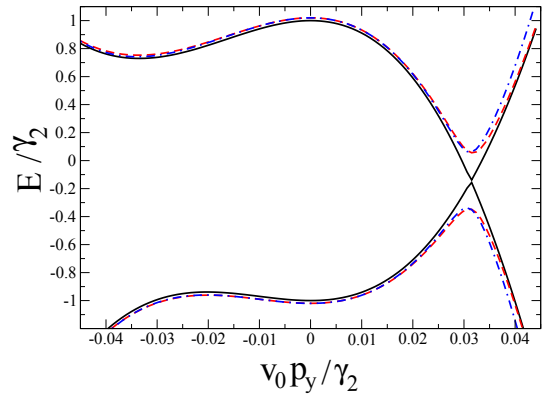}
\caption{Low-energy bands of ABC trilayer graphene at the $K$ point as a function of 
$p_y$ for $p_x=0$: 
using the full $\mathbf{k}\cdot\mathbf{p}$ Hamiltonian with zero spin-orbit 
coupling (solid, black), with finite spin-orbit coupling (red, dashed), and
using the effective Hamiltonian  $\hat{H}^{\rm eff}_{ABC}$ (dashed-dotted, blue).
The parameters $v_i$ were taken from Ref.~\onlinecite{macdonald-1} 
and we have chosen $\lambda_{1/2}=\lambda_1=0.2\,\gamma_2$.}
 \label{fig2}
\end{figure}

In zero external magnetic field  $\hat{H}^{\rm eff}_{ABC}$ is easily diagonalizable. 
Keeping only the leading terms (\ref{so-ham-compact}) in $\hat{H}^{\rm  \,eff }_{so}$, 
we obtain the eigenvalues 
$E_{\pm}=- 2 \frac{v_0 v_4}{\gamma_1} p^2 \pm r(\mathbf{p})$ (each doubly degenerate) where 
$r(\mathbf{p})=\sqrt{\lambda_{1/2}^2+|c(\mathbf{p})|^2+|d(\mathbf{p})}|^2$, 
$c(\mathbf{p}) = -\tilde{\lambda}_{1} \frac{ v_0}{\gamma_1}  p e^{i \phi_{\mathbf{p}}}$, 
$d(\mathbf{p}) = \gamma_2 -\frac{2 v_0 v_3}{\gamma_1} p^2 +\frac{v_0^3}{\gamma_1^2} p^3 e^{- 3 i \phi_{\mathbf{p}}}$ where
$p=|\pi|$) whereas $\phi_{\mathbf{p}}$ is the phase of $\pi$.
The main effect of  SOC on the spectrum is, similarly to monolayer\cite{kane-mele} and bilayer\cite{guinea-so,jfabian-bilayer} graphene, 
to open a band gap  $E_{bg}=2 \sqrt{\lambda_{1/2}^2+|c(\mathbf{p}_{d})|^2}$ % between the bonding and anti-bonding states 
at the band degeneracy points $\mathbf{p}_d$, while preserving the spin degeneracy of the bands (see Fig.~\ref{fig2}). 
Comparing $E_{bg}$ to the SOC band gap
in monolayer and bilayer graphene at the $K$ point, we expect that in ABC trilayer it should be  somewhat bigger 
due to the $|c(\mathbf{p}_{d})|^2$ term, 
i.e. because the band gap can be found away from the $K$ point at finite  $\mathbf{p}_{d}$. 
In Fig.~\ref{fig2} we compare the low-energy bands calculated using the full Hamiltonian (which 
 includes the high-energy bands as well) and using the effective Hamiltonian 
 $\hat{H}^{\rm eff}_{ABC}$ .From this, we conclude that the effective theory
represents a good approximation.
%Indeed, the numerical calculations of Ref.~\onlinecite{konschuh-phd} seem to support  this prediction. 

\section{Substrate induced SOC}
\label{sec:induced-soc}

The calculations of Sect.~\ref{sec:intrinsic-soc} - \ref{sec:SO-ABC-SL} suggest that 
the intrinsic SOC  in ABC trilayer is relatively small, 
the order of magnitude of the SOC parameters is $10\mu{\rm eV}$, the same as in 
monolayer\cite{boettger,jfabian-so-numeric,jfabian-mono-TB,abdelouahed} or bilayer\cite{jfabian-bilayer} graphene. 
One can, in principle, enhance the SOC 
in a number of ways, e.g. by impurities\cite{castro-neto}, 
by applying strong external electric field\cite{jfabian-bilayer}, making 
the graphene sheet curved\cite{klimovaja-curvedbi}, using adatoms with large atomic 
number\cite{graphene-adatom,graphene-adatom-2,graphene-adatom-3} or
bringing the trilayer into proximity with a suitable substrate\cite{proximity-spin-orbit,varykhalov-nature}.
Since proximity to a substrate or adatoms is likely to lead to a larger SOC effect than what one can induce 
by an external electric field, here we consider an effective model whereby strong SOC  is induced in  
one of the outer layers of  trilayer graphene  whereas SOC is not altered in the other two layers. 
For concreteness, we assume that it is the first graphene layer where strong SOC is induced and for 
simplicity we will refer both to  the scenario involving a  substrate and that involving adatoms as  
''substrate induced SOC``.  
From the symmetry point of view,  the model  we consider here is not exact: 
a group theoretical analysis\cite{dresselhaus-book,dresselhaus-paper,jfabian-bilayer} of the matrix elements 
of the  SOC Hamiltonian  shows that by 
breaking the inversion symmetry there can be in principle $21$ different SOC parameters. 
However, we assume that all
intra and interlayer SOC parameters 
%in layers that are not adjacent to the substrate 
will remain small
with respect to the  SOC parameters in the layer that is in the immediate proximity of the
substrate. 
The relevant part of the SOC Hamiltonian that we consider  as a minimal model 
is shown in Table \ref{tbl:induced-SO}.
We assume that at the $K$ point of the BZ only four  SOC parameters have significant values and neglect all other 
intra or inter-layer coupling spin-orbit parameters. 
\begin{table}[htb]
\begin{tabular}{l|ccc}\hline\hline\vspace*{-0.8em}
 & & &\\
SOC  &         $\Psi^{A1}_{\rm eff}$          &          $\Psi^{B3}_{\rm eff}$          &          $\Psi^{B1}_{\rm eff}$   \\ 
\vspace*{-0.8em}
 & & &\\
          \hline \vspace*{-0.8em}
 & & &\\          
\vspace*{-0.8em}      
$\Psi^{A1}_{\rm eff}$   &         $\tilde{\lambda}_{1/2} S_{z}$          &        $0$        &        $\lambda_{BR} S_{+}^{\tau}$    \\ 
 & & &\\        
\vspace*{-0.8em}
$\Psi^{B3}_{\rm eff}$   &        $0$        &       $-\lambda_{1/2} S_{z}$      &        $0$      \\ 
 & & &\\        
\vspace*{-0.8em}      
$\Psi^{B1}_{\rm eff}$    &        ${\lambda}_{BR} S_{-}^{\tau}$        &        $0$        &            $-\tilde{\lambda}_{3/3}^{z} S_{z}$     \\
 & & &\\           
%\vspace*{-0.8em}
\hline\hline
\end{tabular}
\caption{The non-zero part of the model SOC Hamiltonian  $\tau\,\tilde{H}_{so}^{ext}$ %of ABC trilayer graphene for substrate induced SOC
.}
 \label{tbl:induced-SO}
\end{table}
The SOC parameters that we keep are $\tilde{\lambda}_{1/2}$ and $\tilde{\lambda}_{3/3}^{z}$ which may describe enhanced diagonal
SOC on $A1$ and $B1$ type atoms, respectively, whereas $\lambda_{BR}$ is the Bychkov-Rashba\cite{rashba} type 
SOC acting only in the first graphene layer. Since we are going to derive an effective low-energy Hamiltonian, we also keep non-zero
the intrinsic SOC parameter  $\lambda_{1/2}$ on atom $B3$, which, however, might be much smaller than 
$\tilde{\lambda}_{1/2}$ and $\tilde{\lambda}_{3/3}^{z}$.

Similarly to Sect.~\ref{sec:eff_so_ham}, for the full Hamiltonian of the system in the vicinity of the $K$ point we take 
$\tilde{H}_{ABC}=\tilde{H}_{\mathbf{k}\cdotp\mathbf{p}}^{ABC} + \tilde{H}_{so}^{ext}$.  
In the Hamiltonian $\tilde{H}_{\mathbf{k}\cdotp\mathbf{p}}^{ABC}$,  as an additional substrate effect, we also include 
a possible shift $\Delta$ of the on-site energies 
of atoms $A1$ and $B1$  with respect to atoms in the other two layers. (Otherwise 
$\tilde{H}_{\mathbf{k}\cdotp\mathbf{p}}^{ABC}$ is the same as in Section~\ref{sec:eff_so_ham}.) 
To study the low energy physics we again use the Schrieffer-Wolff transformation to eliminate the high-energy bands. 
The hopping  $\gamma_1$ is treated as a large energy scale
with respect to $v_0 \pi $, $|\gamma_3|$, $\gamma_4$, $\gamma_2$, $\Delta$  and we 
keep only the leading order for the terms involving $\Delta$, $\gamma_2$, $v_3$ and $v_4$. 
The electronic part $\hat{\tilde{H}}^{\rm eff}_{el}$ of the effective Hamiltonian $\hat{\tilde{H}}^{\rm eff}_{ABC}$ 
contains one new term in addition to the terms in  Eq.~(\ref{inv_sym_el_eff}):
\begin{equation}
\hat{H}_{\Delta} = \Delta \left[ 
\left(
\begin{array}{cc}
 1 & 0\\
 0 & 0
\end{array}
\right) 
- \frac{v_0^2}{2 \gamma_1 ^2} 
\left(
\begin{array}{cc}
 2\pi^{\dagger}\pi+\pi\pi^{\dagger} & 0\\
 0 & -\pi\pi^{\dagger}
\end{array}
\right)
\right].
\end{equation}

Up to linear order in the momentum and the SOC parameters the effective spin-orbit Hamiltonian has the following terms: 
\begin{subequations}
\begin{eqnarray}
\hat{\tilde{H}}_{so}^{ABC}&=& \left[\hat{\tilde{H}}_{so}^{ABC,(0)}+\hat{\tilde{H}}_{so}^{ABC,(1)}+\hat{\tilde{H}}_{so}^{ABC,(2)}\right],\nonumber\\
\hat{\tilde{H}}_{so}^{ABC,(0)} &=&  
 \tau \,S_z \left(
\begin{array}{cc}
\tilde{\lambda}_{1/2} & 0 \\
0 & -\lambda_{1/2}
\end{array}
\right),  \\
\hat{\tilde{H}}_{so}^{ABC,(1)} &=&  -\tau\,\lambda_{BR}\frac{v_3}{\gamma_1}
\left(
 \begin{array}{cc}
  0  & S_{+}^{\tau} \pi \\
  S_{-}^{\tau} \pi^{\dagger} & 0\\
 \end{array}
 \right), \\
 \hat{\tilde{H}}_{so}^{ABC,(2)} &=&  -\tau\,\lambda_{BR}\frac{v_4}{\gamma_1}
 (S_{-}^{\tau} \pi^{\dagger} + S_{+}^{\tau} \pi)
 \left(
 \begin{array}{cc}
  1  & 0 \\
  0 & 0
 \end{array}
 \right). 
%\label{rashba-so-trilayer}
\end{eqnarray}
\label{rashba-so-trilayer}
\end{subequations}

It is interesting to compare this result to what one obtains for bilayer graphene using the same model\cite{simple-so-bilayer}:
\begin{subequations}
\begin{eqnarray}
\hat{\tilde{H}}_{so}^{AB}&=&\tau\cdotp\left[\hat{\tilde{H}}_{so}^{AB,(0)}+\hat{\tilde{H}}_{so}^{AB,(1)}+\hat{\tilde{H}}_{so}^{AB,(2)}\right],\nonumber\\
\hat{\tilde{H}}_{so}^{AB,(0)} &=&  
 S_z \left(
\begin{array}{cc}
\tilde{\lambda}_{3/3}^{z} & 0 \\
0 & -\lambda_{3/3}^{z}
\end{array}
\right),\\
\hat{\tilde{H}}_{so}^{AB,(1)} &=&  -\lambda_{BR}\frac{v_0}{\gamma_1}
\left(
 \begin{array}{cc}
  0  & S_{+}^{\tau} \pi^{\dagger} \\
  S_{-}^{\tau} \pi  & 0\\
 \end{array}
 \right), \\
 \hat{\tilde{H}}_{so}^{AB,(2)} &=&  -\lambda_{BR}\frac{v_4}{\gamma_1}
 (S_{-}^{\tau} \pi^{\dagger} + S_{+}^{\tau} \pi)
 \left(
 \begin{array}{cc}
  1  & 0 \\
  0 & 0
 \end{array}
 \right). 
\end{eqnarray}
\label{rashba-so-bilayer}
\end{subequations}
In the appropriate limit  Hamiltonian (\ref{rashba-so-bilayer}) 
agrees with the result of Ref.~\onlinecite{jfabian-bilayer}.
Similar models  have also been considered in two recent works:  In Ref.~\onlinecite{mireles} strong Rashba SOC 
was assumed in both layers of bilayer graphene and a linear-in-momentum low-energy SOC Hamiltonian was derived, 
whereas in Ref.~\onlinecite{prada} the authors  
kept  only the first term, $\hat{\tilde{H}}_{so}^{AB,(0)}$  and set  $\lambda_{3/3}^{z}=0$. 

One can see that in contrast to monolayer graphene, in  bilayer and ABC trilayer
graphene the Rashba SOC affects the spin-dynamics in the low-energy bands 
through terms which are momentum dependent\cite{second-order-soi}. This means that at the 
$K$ ($K'$) point the effect of Rashba-type SOC is suppressed with respect to monolayer graphene. 
Furthermore, noting that $v_0\gg v_3$, a comparison of the prefactors of    
$\hat{\tilde{H}}_{so}^{ABC,(1)}$ and $\hat{\tilde{H}}_{so}^{AB,(1)}$  
may suggest that the influence of linear-in-momentum terms on spin-dynamics might be more important in bilayer 
than in trilayer. This is, strictly speaking, only true  in the model that we used, i.e when other  off-diagonal
SOC parameters can be neglected with respect to $\lambda_{BR}$. In general, there would be a 
linear-in-momentum SOC Hamiltonian with a pre-factor proportional to $v_0/\gamma_1$ for the 
trilayer case as well. 
The model introduced above could be relevant e.g. in  
an experiment similar to Ref.~\onlinecite{varykhalov-nature} 
if bilayer or trilayer is used instead of monolayer graphene. 
Varykhalov \emph{et al}\cite{varykhalov-nature} 
reported a large Rashba  SOC in monolayer graphene with $\lambda_{BR}$ between $10-100\, {\rm meV}$,
whereas, as we have seen, the intrinsic SOC  parameters are typically of a few $10 \mu{\rm eV}$.

A detailed study of the properties of  Hamiltonians (\ref{rashba-so-trilayer}) and
(\ref{rashba-so-bilayer}) is left for a future study. We expect a rich physics emerging
from the interplay of diagonal and off-diagonal spin-orbit terms   
and the interlayer asymmetry $\Delta$.

\section{Conclusions}

In conclusion, we studied the intrinsic and substrate induced spin-orbit 
interaction in bilayer and ABC trilayer graphene.
Assuming that in flat graphene systems the most important contribution to the SOC 
comes from the admixture of $p_z$ and $d$ orbitals and using a combination of group-theoretical 
and tight-binding approaches we derived the {intrinsic} SOC Hamiltonian of ABC trilayer graphene. 
In contrast to the similar calculations for monolayer graphene\cite{jfabian-mono-TB}, we found 
that in bilayer and ABC trilayer  in addition to $d_{xz}$ and $d_{yz}$ orbitals 
also $d_{z^2}$ orbitals have to be taken into account. 
For both bilayer and trilayer graphene we obtained explicit expressions for the SOC parameters 
in terms of SK hopping parameters. By comparing these expressions  with the  
DFT calculations of Ref.~\onlinecite{jfabian-bilayer}, 
we were able to estimate the values of the intrinsic SOC constants for ABC trilayer graphene. 
Since the intrinsic SOC is quite small, we considered a situation 
when adatoms or a substrate can induce a strong SOC  (intrinsic diagonal or Rashba type off-diagonal) 
in only one of the  layers of bilayer and ABC trilayer graphene. To describe the low-energy physics we derived 
effective Hamiltonians for both systems. We found that the effect of Rashba type SOC is suppressed close 
to the $K$ ($K'$) point with respect to monolayer graphene. 

{The approach that we used here to derive the SOC Hamiltonians can be employed
in the case of other related problems as well. 
For instance ABA stacked trilayer graphene or graphite can be treated on the same footing 
 when one takes into account,  that they have different symmetries from bilayer and ABC trilayer graphene. 
Considering the substrate induced SOC, which can be strong enough to make the SOC related phenomena experimentally 
observable, one interesting question is  whether the different symmetries 
and band structure of ABC and ABA  trilayer would manifest themselves in e.g. significantly 
different spin-transport properties.}

\section{Acknowledgments}
We acknowledge funding from the DFG within SFB 767 and  from the ESF/DFG within the
EuroGRAPHENE project CONGRAN.

\appendix

\section{Small group of the K point and basis functions}
\label{sec:irrep-basis-funct}

The relevant symmetry group for the calculations at the $K$ point for both trilayer and bilayer graphene is the group  $32$ ($D_3$). 
The generators of this group  are two  three fold rotations ($C_3$) around an axis perpendicular to the plane of the graphene layers 
and three two-fold rotations ($C_2$). 
There are three irreducible representations, two one-dimensional denoted by $\Gamma_{A_1}$, $\Gamma_{A_2}$ and a 
two-dimensional one, $\Gamma_{E}$  (see Table \ref{tbl:D3-group}). 
\begin{table}[hbt]
\begin{tabular}{|c|ccc|}\hline
 $ 32\, (D_3)$ &  $E$  &  $2 C_3$  &  $3 C_2$ \\
 \hline
 $\Gamma_{A_1}$ & $1$  & $1$ & $1$ \\ 
 $\Gamma_{A_2}$ & $1$  & $1$ & $-1$ \\
 $\Gamma_{E}$ & $2$  & $-1$ & $0$\\
 \hline\hline
\end{tabular}
\caption{Character table of the group $32$ ($D_3$).}
\label{tbl:D3-group}
\end{table}

The basis functions built from the linear combinations of $p_z$ and $d$ orbitals
which transform as these irreducible representations are shown in Table \ref{tbl:symmetry-basis-full-trilayer} for 
ABC trilayer and in Table \ref{tbl:symmetry-basis-full-bilayer} for bilayer. 
\begin{table}[hbt]
\begin{tabular}{|c|c|}\hline\vspace*{-0.8em}
 & \\
\vspace*{-0.8em}
 $\Gamma_{A_1}$ &  $|\overline{\Psi_{1,0,0}^{1,3}}\rangle$, $|\Psi_{2,1,-1}^{3,1}\rangle$, $|\Psi_{2,0,0}^{1,3}\rangle$, 
$|\Psi_{2,-2,2}^{3,1}\rangle$, $|\Psi_{2,-1,1}^{2,2}\rangle$, $|\Psi_{2,2,-2}^{2,2}\rangle$\\
 & \\  
\hline\vspace*{-0.8em} 
 & \\
\vspace*{-0.8em}
 $\Gamma_{A_2}$ & $|{\Psi_{1,0,0}^{1,3}}\rangle$,  $|\overline{\Psi_{2,1,-1}^{3,1}}\rangle$,
  $|\overline{\Psi_{2,0,0}^{1,3}}\rangle$, $|\overline{\Psi_{2,-2,2}^{3,1}}\rangle$, 
  $|\overline{\Psi_{2,-1,1}^{2,2}}\rangle$, $|\overline{\Psi_{2,2,-2}^{2,2}}\rangle$ \\ 
 & \\  
  \hline\vspace*{-0.8em}
 & \\
%\vspace*{-0.8em}
 $\Gamma_{E}$ & $\{|{\Psi_{1,0,0}^{2,1}}\rangle, |{\Psi_{1,0,0}^{3,2}}\rangle\}$,  
                $\{|\overline{\Psi_{1,0,0}^{2,1}}\rangle, |\overline{\Psi_{1,0,0}^{3,2}}\rangle\}$ \\
\vspace*{-1.0em}
 & \\
              &  $\{|{\Psi_{2,0}^{A2}}\rangle, |{\Psi_{2,0}^{B2}}\rangle\}$, 
                 $\{|{\Psi_{2,1}^{A2}}\rangle, |{\Psi_{2,-1}^{B2}}\rangle\}$,
                 $\{|{\Psi_{2,-2}^{A2}}\rangle, |{\Psi_{2,2}^{B2}}\rangle\}$ \\
\vspace*{-1.0em}
 & \\
              &   $\{|{\Psi_{2,0}^{A3}}\rangle, |{\Psi_{2,0}^{B1}}\rangle\}$, 
                  $\{|{\Psi_{2,-1}^{A3}}\rangle, |{\Psi_{2,1}^{B1}}\rangle\}$, 
                  $\{|{\Psi_{2,2}^{A3}}\rangle, |{\Psi_{2,-2}^{B1}}\rangle\}$\\
\vspace*{-1.0em}
 & \\
              &   $\{|{\Psi_{2,1}^{A1}}\rangle, |{\Psi_{2,-1}^{B3}}\rangle\}$, 
                  $\{|{\Psi_{2,-1}^{A1}}\rangle, |{\Psi_{2,1}^{B3}}\rangle\}$,
                  $\{|{\Psi_{2,2}^{A1}}\rangle, |{\Psi_{2,-2}^{B3}}\rangle\}$\\
\vspace*{-1.0em}
 &\\
              &   $\{|{\Psi_{2,-2}^{A1}}\rangle, |{\Psi_{2,2}^{B3}}\rangle\}$\\
%\vspace*{-0.8em}              
 & \\
%\vspace*{-1.0em}
\hline\hline
\end{tabular}
\caption{Basis functions for the irreducible representations of the small group of the $K$ point for ABC trilayer graphene. 
The basis functions for the $K'$ point can be obtained by complex-conjugation.}
\label{tbl:symmetry-basis-full-trilayer}
\end{table}

\begin{table}[hbt]
\begin{tabular}{|c|c|}\hline\vspace*{-0.8em}
 & \\
\vspace*{-0.8em}
 $\Gamma_{A_1}$ &  $|\overline{\Psi_{1,0,0}^{2,1}}\rangle$, $|\Psi_{2,1,-1}^{1,2}\rangle$, $|\Psi_{2,0,0}^{2,1}\rangle$, 
$|\Psi_{2,-2,2}^{2,1}\rangle$ \\
 & \\  
\hline\vspace*{-0.8em} 
 & \\
\vspace*{-0.8em}
 $\Gamma_{A_2}$ & $|{\Psi_{1,0,0}^{2,1}}\rangle$,  $|\overline{\Psi_{2,1,-1}^{1,2}}\rangle$,
  $|\overline{\Psi_{2,0,0}^{2,1}}\rangle$, $|\overline{\Psi_{2,-2,2}^{2,1}}\rangle$ \\ 
 & \\  
  \hline\vspace*{-0.8em}
 & \\
%\vspace*{-0.8em}
 $\Gamma_{E}$  &  $\{|{\Psi_{1,0}^{A1}}\rangle, |{\Psi_{1,0}^{B2}}\rangle\}$, 
                 $\{|{\Psi_{2,0}^{A1}}\rangle, |{\Psi_{2,0}^{B2}}\rangle\}$,
                 $\{|{\Psi_{2,-1}^{A1}}\rangle, |{\Psi_{2,1}^{B2}}\rangle\}$ \\
\vspace*{-1.0em}
 & \\
              &   $\{|{\Psi_{2,2}^{A1}}\rangle, |{\Psi_{2,-2}^{B2}}\rangle\}$, 
                  $\{|{\Psi_{2,-2}^{A1}}\rangle, |{\Psi_{2,2}^{B1}}\rangle\}$, 
                  $\{|{\Psi_{2,-1}^{A2}}\rangle, |{\Psi_{2,1}^{B1}}\rangle\}$\\
\vspace*{-1.0em}
 & \\
              &   $\{|{\Psi_{2,1}^{A2}}\rangle, |{\Psi_{2,-1}^{B1}}\rangle\}$, 
                  $\{|{\Psi_{2,2}^{A2}}\rangle, |{\Psi_{2,-2}^{B1}}\rangle\}$\\
%\vspace*{-0.8em}              
 & \\
%\vspace*{-1.0em}
\hline\hline
\end{tabular}
\caption{The same as in Table \ref{tbl:symmetry-basis-full-trilayer} but for bilayer graphene.}
\label{tbl:symmetry-basis-full-bilayer}
\end{table}

In the case of bilayer, making then the same steps as for trilayer graphene, 
one finds  the symmetry basis and intrinsic SOC parameters given in Table \ref{tbl:SlatKost-SOC-bilayer}.
\begin{table*}[htb]
\begin{tabular}{l|l}\hline\hline\vspace*{-0.8em}
 & \\
$|\Psi_{\Gamma_{A_{1}}}\rangle=|\overline{\Psi_{1,0,0}^{2,1}}\rangle+
              \frac{3}{\sqrt{2}}\frac{V_{p d \pi}}{\delta\vareps_{pd}-\gamma_1} |\Psi_{2,1,-1}^{1,2}\rangle 
           -\frac{{V}_{p d \sigma}}{\delta\vareps_{pd}-\tilde{\gamma}_1} |\Psi_{2,0,0}^{2,1}\rangle$           &
\,\, $\lambda_{1/2}=-\frac{9}{2} \,\xi_d\, \frac{(V_{p d\pi})^2}{(\delta\vareps_{pd}^2-\gamma_1^2)}$ \\
\vspace*{-0.8em}
 & \\
\vspace*{-0.8em}
$|\Psi_{\Gamma_{A_{2}}}\rangle=|\Psi_{1,0,0}^{2,1}\rangle-
           \frac{3}{\sqrt{2}}\frac{V_{p d \pi}}{\delta\vareps_{pd}+\gamma_1} |\overline{\Psi_{2,1,-1}^{1,2}}\rangle 
           +\frac{{V}_{p d \sigma}}{\delta\vareps_{pd}+\tilde{\gamma}_1} |\overline{\Psi_{2,0,0}^{2,1}}\rangle$    &   
\,\, $\lambda_{1/3} =- 3 \sqrt{\frac{3}{2}}\, \xi_d\, 
                    \frac{V_{p d \pi} {V}_{p d \sigma}}{\delta\vareps_{pd} (\delta\vareps_{pd}-\tilde{\gamma}_1)}$ \\         
 & \\              
\vspace*{-0.8em}
$|\Psi_{\Gamma_{E_{1,1}}}\rangle=|\Psi_{1,0}^{A1}\rangle+ 
                                  \frac{3}{\sqrt{2}}\frac{V_{p d \pi}}{\delta\vareps_{pd}} |\Psi_{2,-1}^{B1}\rangle$
              &   
\,\, $\lambda_{2/3} =- 3 \sqrt{\frac{3}{2}}\, \xi_d\, 
                    \frac{V_{p d \pi} {V}_{p d \sigma}}{\delta\vareps_{pd} (\delta\vareps_{pd}+\tilde{\gamma}_1)}$\\
 & \\                   
\vspace*{-0.8em}
$|\Psi_{\Gamma_{E_{1,2}}}\rangle=|\Psi_{1,0}^{B2}\rangle-
                                \frac{3}{\sqrt{2}}\frac{V_{p d \pi}}{\delta\vareps_{pd}} |\Psi_{2,1}^{A2}\rangle$  &  
\,\, $\lambda_{3/3}^{z}=-\frac{9}{2} \,\xi_d\, \frac{(V_{p d\pi})^2}{\delta\vareps_{pd}^2}$  \\
& \\
\hline\hline
\end{tabular}
\caption{Symmetry basis functions (left column) and intrinsic spin-orbit matrix elements (right column) in terms of  
Slater-Koster parameters for bilayer graphene. We used the notation  $V_{p d \sigma}=V_{p d \sigma}^{A2,B1}$, 
$\delta\vareps_{pd}=\vareps_p-\vareps_d$ and $\tilde{\gamma}_1=\gamma_1+V_{d d \sigma}$.}
\label{tbl:SlatKost-SOC-bilayer}
\end{table*}
Referring to  Fig.~\ref{fig1}(a),  the atomic sites participating in  the formation of  
the split-off bands are denoted by  $B1$ and $A2$, whereas they are labeled as
$A1$ and $B2$ in Ref.~\onlinecite{jfabian-bilayer}, and similarly for the sites
contributing to the low energy bands. Therefore, to arrive at the SOC Hamiltonian  shown in Table \ref{tbl:intrinsic-SO-bilayer}, 
(i) one has to rotate the symmetry basis  into the on-site basis,
(ii) re-label the sites as $Aj \leftrightarrows Bj$, (iii) and make the identification $\lambda_{\rm I 1}=-\lambda_{3/3}^z$, 
$\lambda_{\rm I 2}=-\lambda_{1/2}$, $\lambda_0=\frac{1}{\sqrt{2}}(\lambda_{2/3}-\lambda_{1/3})$, 
$\lambda_4^{bi}=-\frac{1}{\sqrt{2}}(\lambda_{2/3}+\lambda_{1/3})$.
The matrix elements are real numbers in our calculations  because we use different lattice vectors   
than in Ref.~\onlinecite{jfabian-bilayer}.

\section{}
Here  we collect some useful formulae:  We give explicitly the transformation
between the symmetry basis and the effective on-site $p_z$ orbital basis for the trilayer case and 
present the low-energy electronic Hamiltonian for bilayer graphene.

\subsection{Transformation between the symmetry basis and the on-site effective $p_z$ basis}
\label{sec:on-site-transf}

The transformation reads:
\begin{equation}
\left(
\begin{array}{c}
 |\Psi^{A1}_{\rm eff}\rangle \\ |\Psi^{B3}_{\rm eff}\rangle\\ |\Psi^{B1}_{\rm eff}\rangle\\   
 |\Psi^{A2}_{\rm eff}\rangle \\ |\Psi^{B2}_{\rm eff}\rangle \\ |\Psi^{A3}_{\rm eff}\rangle
\end{array}
\right)
=\frac{1}{\sqrt{2}}
\left(
\begin{array}{cccccc}
  1 & 1 & 0 & 0 & 0 & 0\\
  -1 & 1 & 0 & 0 & 0 & 0 \\
  0 & 0 & 1 & 0 & -1& 0 \\
  0 & 0 & 1 & 0 & 1& 0 \\
  0 & 0 & 0 & 1 & 0 & -1 \\
  0 & 0 & 0 & 1 & 0 &  1 \\
\end{array}
\right)
\left(
\begin{array}{c}
|\Psi_{\Gamma_{A_1}}\rangle\\  |\Psi_{\Gamma_{A_2}}\rangle \\ |\Psi_{\Gamma_{E_{1,1}}}\rangle\\
|\Psi_{\Gamma_{E_{1,2}}}\rangle\\ |\Psi_{\Gamma_{E_{2,1}}}\rangle \\ |\Psi_{\Gamma_{E_{2,2}}}\rangle  
\end{array}
\right).
\end{equation}

\subsection{Bilayer graphene low-energy electronic Hamiltonian}

For completeness and for comparison with the trilayer case given in Eq.~(\ref{inv_sym_el_eff}), 
we show here the low-energy electronic Hamiltonian of bilayer graphene in the on-site basis $A1$, $B2$. This Hamiltonian 
has been discussed in many publications before, see e.g. the recent review of Ref.~\onlinecite{mccann-koshino-bilayer}.
As in Sect.~\ref{sec:induced-soc}, we assume that atoms $A1$ and $B1$ in the layer adjacent to the substrate 
have a different on-site energy $\Delta$ than atoms $A2$, $B2$ in the second layer. 
The most important terms are found to be:
\begin{eqnarray}
{\hat{H}}^{\rm \, eff }_{el} &=& {\hat{H}}_{chir} + {\hat{H}}_{3w} + {\hat{H}}_{v_4}+\hat{H}_{\Delta} , \nonumber \\
{\hat{H}}_{chir} &=& -\frac{v_0^2}{\gamma_1}\left(
\begin{array}{cc}
0 & \left( {\pi }^{\dag }\right)^{2} \\
{\pi^{2}} & 0
\end{array} \right),\,\,\,
\hat{H}_{3w} =
v_3
 \left(
\begin{array}{cc}
0 & \pi \\
\pi^{\dagger} & 0
\end{array}
\right),\nonumber \\
{\hat{H}}_{v_4} & = &
-\frac{2 v_0 v_4}{\gamma_1}
\left(
\begin{array}{cc}
\pi^{\dagger} \pi & 0 \\
0 & \pi \pi^{\dagger}
\end{array}
\right),\nonumber \\
\hat{H}_{\Delta} &=&\Delta \left[1-\frac{v_0^2}{2 \gamma_1^2} \pi^{\dagger}\pi\right]
\left(
\begin{array}{cc}
 ×1 & 0\\
  0 & 0
\end{array}
\right).
\label{}
\end{eqnarray}

\bibliographystyle{prsty}

\begin{thebibliography}{10}
\bibitem{zeros}
Zero matrix elements can arise e.g. because we neglect remote direct hoppings 
between the first and the third layer except for the hopping $A1$-$B3$.
\bibitem{S-approx}
The approximation $e^{-S}\approx 1-S $ can be justified by the numerical calculations of 
Ref.~\onlinecite{jfabian-mono-TB}.  There it was shown that in the case of monolayer
graphene $\frac{3}{2} \frac{V_{p d \pi}}{(\vareps_p-\vareps_d)} \approx 0.0871$. Since 
$\gamma_2$ is a small energy scale compared to $\delta\vareps_{pd}=\vareps_p-\vareps_d$, 
we find that $\tilde{V}_{p d \pi}/(\delta\vareps_{pd}-\gamma_2)\ll 1$. 
Furthermore, since $V_{p d \sigma}^{A1, B3}$ and $V_{d d \sigma}^{A1,B3} $ correspond
to hopping between $A1$ and $B3$ atoms 
which are at larger distance than the $A$ and $B$ atoms within the same graphene layer, one 
can expect that $V_{p d \sigma}^{A1,B3} < V_{p d \pi}$ and $V_{d d \sigma}^{A1,B3} \propto \gamma_2$, 
hence both matrix elements of  $S^{(1)}_{pd}$ in Eq.~(\ref{S-approx}) are much smaller than unity.
\bibitem{two-soc-param}
Ref.~\onlinecite{guinea-so} considers an effective one-layer model and obtains a SOC Hamiltonian which 
is wavenumber independent and can be characterized by two SOC constants. One of the SO parameters, 
denoted by $\lambda_1$ in Ref.~\onlinecite{guinea-so}, corresponds to our $\lambda_{1/2}$. 
This can be seen by folding down the full Hamiltonian $\hat{H}_{ABC}$ to obtain a low-energy 
effective model [see Eq.~(\ref{so-ham-compact}) in Sect.~\ref{sec:eff_so_ham}]. 
Further wavenumber-independent  SOC terms 
can also be obtained from the folding down procedure, but they are higher order in the 
SOC parameters. Therefore we think that they are negligible with respect to 
the effective SOC Hamiltonian shown  in Eq.~(\ref{so-ham-compact}). 

Ref.~[\onlinecite{konschuh-phd}] explains the numerically calculated band structure with a single, 
diagonal SOC parameter.
\bibitem{simple-so-bilayer}
In general, if the inversion symmetry is broken there are ten SOC parameters in bilayer graphene. 
As in Ref.~\onlinecite{jfabian-bilayer},  we keep only the most important ones.
\bibitem{second-order-soi}
More precisely, as we have already mentioned in connection with 
the intrinsic SOC\cite{two-soc-param}, when one folds down the full Hamiltonian 
one can obtain  additional momentum-independent terms for the low-energy Hamiltonian 
but they are proportional to products and higher powers of the SOC parameters. 
We neglect these terms.
\end{thebibliography}

\end{document}